\documentclass[prd, amsfonts, twocolumn, nofootinbib, showpacs]{revtex4}

\usepackage{graphicx, epsfig}
\usepackage{color}
\usepackage{amsmath}
\usepackage{mathtools}
\newcommand{\be}{\begin{equation}}
\newcommand{\ee}{\end{equation}}
\newcommand{\bea}{\begin{eqnarray}}
\newcommand{\eea}{\end{eqnarray}}

\newcommand{\gapp}{\mathrel{\raise.3ex\hbox{$>$}\mkern-14mu \lower0.6ex\hbox{$\sim$}}}
\newcommand{\lapp}{\mathrel{\raise.3ex\hbox{$<$}\mkern-14mu \lower0.6ex\hbox{$\sim$}}}
\newcommand{\LSM}{L$\Sigma$M}

\newcommand{\GMLfull}{Gell-Mann-L\'evy~}

\newcommand{\HVEV}{\langle H\rangle}
\newcommand{\SVEV}{\langle S\rangle}
\newcommand{\half}{\frac{1}{2}}
\newcommand{\mpisq}{m_\pi^2}

\newcommand{\GlnuSM}{{\nu}{\mathrm {SM}}^{\rm G}_{ude\nu}}
\newcommand{\GlnuDSM}{{\nu_D}{\mathrm {SM}}^{\rm G}_{tb\tau\nu_\tau}}

\def\bbox{{\,\lower0.9pt\vbox{\hrule \hbox{\vrule height 0.2 cm
\hskip 0.2 cm \vrule  height 0.2 cm}\hrule}\,}}

\begin{document} 

\title{Global $SU(3)_C\times SU(2)_L\times U(1)_Y$
linear sigma  model $\GlnuDSM$:
\\axial-vector Ward Takahashi identities, and
decoupling of certain heavy \\BSM particles
due to the Goldstone theorem}

\author{Bryan W. Lynn$^{1,2,3}$ and Glenn D. Starkman$^{1}$ 
}
\affiliation{$^1$ ISO/CERCA/Department of Physics, Case Western Reserve University, Cleveland, OH 44106-7079}
\affiliation{$^2$ University College London, London WC1E 6BT, UK}
\affiliation{$^3$ Department of Physics, University of Wisconsin, Madison, WI 53706-1390}
\email{bryan.lynn@cern.ch,  gds6@case.edu}

\begin{abstract}
{\bf This work is dedicated to the memory of Raymond Stora (1930-2015).}
 In the  $SU(2)_L\times SU(2)_R$ Linear Sigma Model with PCAC,
a tower of Ward-Takahashi Identities (WTI) have long been known 
to give relations among 
 1-{\em Scalar}-Particle-Irreducible (1-$\phi$-I) Green's functions, 
 and  among I-{\em Scalar}-Particle-Reducible (1-$\phi$-R) T-Matrix elements for external scalars 
(i.e. the Brout-Englert-Higgs (BEH) scalar $H$, and 3 pseudoscalars ${\vec \pi}$). 
In this paper, we extend these WTI and the resulting relations 
to the $SU(3)_C\times SU(2)_L\times U(1)_Y$ Linear Sigma Model including the heaviest generation of 
Standard Model (SM) fermions -- the ungauged (i.e. global) Standard Model SM$^G_{tb\tau\nu_\tau}$ --
supplemented with the minimum necessary neutrino content -- 
right-handed neutrinos and Yukawa-coupling-induced Dirac neutrino mass, 
to obtain the $CP$-conserving $\GlnuDSM$,
and extract powerful constraints on the effective  Lagrangian: 
e.g. showing that they make separate tadpole renormalization unnecessary, and guarrantee infra-red finiteness.
The crucial observation is that ultra-violet quadratic divergences (UVQD),
and all other relevant operators, 
contribute {\bf only} to $\mpisq$, 
a {\it pseudo}-Nambu-Goldstone boson (NGB) mass-squared, 
which appears in intermediate steps of calculations.
A WTI between Transition-Matrix elements 
(or, in this global theory equivalently the Goldstone Theorem) 
then enforces $\mpisq=0$ exactly for the true NGB
in the spontaneous symmetry breaking (SSB) mode of the theory. 
{\bf The Goldstone Theorem thus causes all relevant operator  contributions}, 
originating to all-loop-orders from virtual scalars $H,{\vec \pi}$, quarks $q_{L}^c;t_{R}^c;b_{R}^c$  and leptons $l_{L};{\nu_\tau}_R;\tau_{R}$ with ($c=r,w,b$),
{\bf to vanish identically!}

We show that our regularization-scheme-independent, WTI-driven results 
are unchanged by the addition of certain $SU(3)_C\times SU(2)_L \times U(1)_Y$ heavy ($M_{Heavy}^2 \gg |q^2|, m_{Weak}^2$) 
$CP$-conserving matter, 
such as originate in certain Beyond the SM (BSM) models. 
The  global axial-vector WTI 
again cause all UVQD and finite relevant operators 
to vanish, in the $\nu_D {\mathrm {SM}}^{\rm G}_{tb\tau\nu_\tau + Heavy}$ model. 
We demonstrate this with two examples:
a singlet $M^2_S \gg m_{Weak}^2$ real scalar field $S$ with 
discrete $Z_2$ symmetry and $\SVEV=0$; and
a  singlet right-handed Type I See-saw Majorana neutrino $\nu_R$ with $M_{\nu_R}^2\gg m_{Weak}^2$.
Specifically, we prove that these heavy degrees of freedom decouple completely from the  low-energy $\GlnuDSM$ effective Lagrangian, contributing only irrelevant operators after quartic-coupling renormalization.  
\end{abstract}

\maketitle

\section{Introduction}

Ward-Takahashi Identities (WTI) are 
relations among Green's functions or amplitudes of field theories 
that result from the symmetries of the theory.  
They exist both in \lq\lq{unbroken}\rq\rq~ theories 
(in which the vacuum shares the symmetries of the Lagrangian) 
and in spontaneously broken theories 
(in which the vacuum does not share the symmetries of the Lagrangian).
In this paper we are concerned specifically with 
the global $SU(2)_L\times U(1)_Y$ Schwinger \cite{Schwinger1957} 
Linear Sigma Model (\LSM),
the ungauged scalar sector of the Standard Model (SM),
augmented by the third generation of SM fermions with their usual Yukawa couplings to the Higgs doublet
(and so augmenting the symmetry with a global $SU(3)_C$ factor),
as well as by a right-handed $\tau$ neutrino, with its allowed Yukawa couplings.
For brevity, we call this global theory the $\GlnuDSM$, 
with G for global and D indicating that the neutrinos have only Dirac masses.
With SM isospin and hypercharge quantum numbers for fermions, 
the 3rd-generation $\GlnuDSM$ has zero axial anomaly. 
We prove here that the $CP$-conserving $\GlnuDSM$ is governed by axial-vector WTI
directly analogous with those  proved by B.W. Lee \cite{Lee1970} 
for the $SU(2)_L\times SU(2)_R$ \GMLfull \cite{GellMannLevy1960} 
with Partially Conserved Axial-vector Currents (PCAC).
One of those axial-vector WTIs is equivalent in this global theory
to the Goldstone Theorem, which protects
the mass of the Nambu-Goldstone bosons (NGB) from non-zero contributions.
\footnote{ 
	In June 2011 \cite{Lynn2011}  one of us (BWL) introduced these ideas.
	A December 2011 pedagogical companion paper \cite{Lynnetal2012} 
	simplified the treatment of UVQDs in the context of 
	the {\em global} \GMLfull model \cite{GellMannLevy1960} with PCAC. 
	In \cite{LynnStarkman2013} we showed that, what we called the Goldstone theorem,
	but to be specific is really a WTI equivalent to the Goldstone Theorem 
	in this global theory,  
	protects the weak-scale {\em global} SSB $SO(2)$ 
		Schwinger model \cite{Schwinger1957} (i.e.
	against 1-loop relevant operators $\sim M_{Heavy}^2 \gg m_{Weak}^2$ 
	which arise from virtual heavy particles) by way of 2 explicit 1-loop examples:
	a real singlet scalar $S$ and a singlet Majorana neutrino $\nu_R$ with $M_S^2,M_{\nu_R}^2 \gg \vert q^2\vert, \HVEV^2$.
	}

We also demonstrate that there exists a wide class 
of heavy matter $M_{Heavy}^2 \gg m_{Weak}^2$ particles 
from which the low-energy effective $\GlnuDSM$ Lagrangian,
fortified by the WTI, 
is protected. 
It may be no coincidence that this class includes 
heavy Majorana masses for right-handed neutrinos, 
as envisioned in the see-saw models of light neutrinos.
Another theory might well have been less effectively protective.

Here we prove properties  of the  spontaneously broken mode of a quantum field theory with global symmetries that are the rigid versions of the local
symmetries of the Standard Model, in anticipation of extending our arguments
to the one-3rd-generation standard electroweak model itself \cite{LSS-4Proof,SU(2)Proof}. 
We discover how the physics of the theory
(as embodied in on-shell Transition-matrix elements) 
is more symmetric than the effective Lagrangian, 
because consistency conditions on the states 
constrain the physics. 
A particularly crucial role is played by a WTI among T-matrix elements,
which is equivalent to the Goldstone Theorem in this global theory.
In upcoming papers we extend these results, 
first to a $U(1)_Y$ gauge theory, the  $CP$-conserving Abelian Higgs model (AHM) \cite{LSS-3Proof},
and then to the $CP$-conserving gauged electroweak SM, with the 3rd generation of quarks, charged leptons, $\nu_L,\nu_R$, and Dirac-mass neutrinos, $\nu_D {\mathrm {SM}}_{tb\tau\nu_\tau}$ \cite{SU(2)Proof,LSS-4Proof}.
Along the way we discover that the important T-matrix WTI and the Goldstone Theorem
contain independent information.

The structure of the remainder of this paper is as follows.

Section \ref{RenormalizationSM} concerns 
the correct (i.e. axial-vector-WTI-obedient) 
renormalization of the scalar-sector effective 
$\GlnuDSM$ Lagrangian in its Goldstone (i.e. spontaneously broken) mode.
In this section  we treat the $\GlnuDSM$,
with its SM fermions, 
augmented by a right-handed neutrino, and consequently a Dirac neutrino mass,  
as a {\bf stand-alone} flat-space quantum field theory,
not embedded or integrated into any higher-scale ``Beyond the SM" (BSM)  physics. 

Section \ref{RenormalizationBSM} extends our results 
to $M_{Heavy}^2 \gg m_{BEH}^2$ heavy $SU(3)_{color}\times SU(2)_L \times U(1)_Y$ matter representations, such as arise in certain BSM models:

Section \ref{Conclusions} draws  a historical lesson.

Appendix \ref{WTIproof} extends the proof of B.W. Lee 
(i.e. for the WTI of $SU(2)_L\times SU(2)_R$ \GMLfull with PCAC),
to WTI for  $SU(3)_C\times SU(2)_L\times U(1)_Y$ $\GlnuDSM$ -- 
the mathematical basis on which the results of this paper rest.

\section{Axial-vector-WTI-obedient renormalization 
of the global $SU(3)_C\times SU(2)_L\times U(1)_Y$ $\GlnuDSM$ effective Lagrangian}
\label{RenormalizationSM}

The global $SU(3)_{C}\times SU(2)_L\times U(1)_Y$ Lagrangian of 
SM scalar and 3rd generation fermion fields
\footnote{	
	The $\nu_D$SM matter fields are well known: 
	a spin $S=0$ complex scalar doublet $\phi$; 
	$S=\half$ left-handed and right-handed leptons 
	$l_{L}^i=\left[\nu^{i}_L e^{i}_L\right]^T ,e^{i}_R$, $\nu^i_R$,
	with $i$ running over the 3 generations;
	$S=\half$ left-handed and right-handed quarks  
	$q^{i,c}_L=\left[d^{i,c}_L u^{i.c}_L\right]^T ,d^{i,c}_R, u^{i,c}_R$, 
	with $i$ running over the 3 generations, 
	and $c=r,w,b$ over the $SU(3)_C$ color index. 
Quarks, and separately leptons, have complex Yukawas, Dirac masses and mixings. 
	The observable $3\times 3$ Cabibbo-Kobayashi-Maskawa (CKM) and Pontecorvo$-$Maki$-$Nakagawa$-$Sakata (PMNS) matrices connect the weak eigenstates with mass eigenstates.
In order to assure CP conservation, we limit ourselves to one generation of fermions,
the third-generation of the Standard Model.
},
extended with a right-handed neutrino with Dirac mass 
\footnote{
	Before the experimental observation of neutrino mixing, 
	the SM was defined to include only left-handed neutrinos $\nu_L^i$.
	The proof of our new  axial-vector WTI's 
	{\em requires} $CP$-conservation and that neutrinos be massive.  
	We therefore pay homage to experimentally observed neutrino mixing, 
	and study in this paper the $CP$-conserving $\GlnuDSM$,
	here defined to include a right-handed neutrino $\nu_R$; and a Dirac mass
	$m^{\nu}_{Dirac} = y_\nu \HVEV /{\sqrt 2}$. 
	}
is
\bea
\label{SMFieldsLagrangian}
&&L_{\nu_DSM^G_{tb\tau\nu_\tau}}\Big(\phi ;l_L, \tau_R , {\nu_\tau}_R;q^{c}_L,b^{c}_R, t^{c}_R;\nonumber \\
&&\qquad \qquad \mu_{\phi}^2, \lambda_{\phi}^2; y_{b}, y_{t}, y_{\tau}, y_{\nu_\tau} \Big) \qquad 
\eea
$\GlnuDSM$ parameters include
quadratic and quartic scalar couplings $\mu _{\phi}^2, \lambda _{\phi}^2$,
\footnote{
	We follow the language and power counting of the early literature \cite{Lee1970}, 
	taking the quartic coupling constant to be $\lambda_\phi^2$
	rather than the modern \cite{Ramond2004} $\lambda$.
	Renormalized  $\lambda^2_{\phi} \ge 0$.
}
and real Yukawa couplings. This Lagrangian conserves $CP$.

We define a complex  BEH doublet representation for the scalars
\begin{equation}
	\label{SMHiggs}
	\phi \equiv  \frac{1}{\sqrt{2}} \left[ \begin{array}{c}H+i\pi_3\\ -\pi_2 + i\pi_1\end{array}\right]\,.  
\end{equation}
(This can trivially be mapped to an O(4) quartet of real scalars, since
$\phi^\dagger\phi = \half (H^2 + \vec{\pi}^2)$.)
We use this manifestly renormalizable linear representation 
for the scalars in order to control relevant operators. 

The Lagrangian (\ref{SMFieldsLagrangian}) has 3 modes, which
we characterize by the values of the renormalized BEH-VEV 
$\HVEV$
and the renormalized ({\it pseudo}-)NGB mass-squared $\mpisq$:
$\HVEV=0,\mpisq > 0$, known as the ``Wigner mode"; 
$\HVEV=\mpisq = 0$, the classically scale-invariant point;  and
$\HVEV \neq 0,\mpisq = 0$, the spontaneously broken or ``Goldstone mode."

This paper distinguishes carefully between 
the global $SU(3)_{color}\times SU(2)_L\times U(1)_Y$ 
Lagrangian of (a single generation of) SM matter fields (i.e. (\ref{SMFieldsLagrangian}))
and the $\GlnuDSM$ itself: i.e. 
the $\GlnuDSM$ is the ``Goldstone mode" of (\ref{SMFieldsLagrangian}).

\medskip
{\bf 1) Symmetric $\HVEV=0,\mpisq \neq 0$ Wigner mode: }\newline
This is analogous with the Schwinger-model x-axis Figure 12-12 
in the textbook by C.Itzykson \& J-C. Zuber \cite{ItzyksonZuber} 
and the similar Figure 1 in \cite{Lynnetal2012}.  
The analysis and renormalization of Wigner mode 
(e.g. its infra-red structure with massless fermions) 
is outside the scope of this paper. 
Thankfully, Nature is not in Wigner mode! 
Our universe is, instead, in the SSB Goldstone mode.

\medskip
{\bf 2) Classically scale-invariant point $\HVEV=0,\mpisq=0$}
 is analogous with the Schwinger-model origin in Figure 12-12 
 in the textbook by C.Itzykson \& J-C. Zuber \cite{ItzyksonZuber} 
 and Figure 1 in \cite{Lynnetal2012}.
The analysis and renormalization of the classically scale-invariant point 
is outside the scope of this paper. 

\medskip
{\bf 3) Spontaneously broken $\HVEV\neq0,\mpisq = 0$ Goldstone mode: }\newline
The $\GlnuDSM$ is the Goldstone mode of $L_{\nu_dSM^G}$ in (\ref{SMFieldsLagrangian}). 
The ``physics" is the spectrum of physical particles -- 
$S=0$ bosons, and the 3rd generation of $S=\half$ Dirac-massive quarks and leptons -- 
and their associated dynamics.

The { one-generation} {\bf global} $\GlnuDSM$ 
is invariant under global $SU(3)_{C}\times SU(2)_L\times U(1)_Y$ transformations 
and conserves $CP$.
Global {\it axial-vector} WTI therefore
govern the dynamics of the $\phi$-sector 
(i.e. BEH scalar $H$ and 3 pseudoscalars ${\vec \pi}$)
of the  all-loop-orders renormalized $\GlnuDSM$ effective Lagrangian%
\footnote{
{ Our interest in  axial-vector WTI may be surprising, 
given that while $SU(2)_L$ is a symmetry of the Lagrangian, $SU(2)_{L-R}$ is not.
This interest is justified by our insistence  on CP conservation,
as described in detail in Appendix \ref{WTIproof}.
In future work, we will consider the interesting consequences
for our WTI, and for physics, of small amounts of CP violation. }	
}.
There are two sets of such $CP$-conserving axial-vector WTI.
One set governs relations 
among connected amputated 1-($h,{\vec \pi}$)ScalarParticle-Irreducible (1-$\phi$-I) 
 Greens functions. 
A separate set  governs relations 
among connected amputated 1-($h,{\vec \pi}$)ScalarParticle-Rreducible (1-$\phi$-R) 
 T-Matrix elements. 
As observed by Lee for the \GMLfull model  
 \footnote{
	Ref. \cite{Lynnetal2012} used B.W. Lee's \GMLfull (GML) WTI 
	to construct the all-loop-orders renormalized low-energy 
	$(\vert q^2\vert, \HVEV^2, m_{Weak}^2 \ll$ Euclidean UV cut-off $\Lambda^2)$,
	effective GML Lagrangian including UV quadratic divergances (UVQD).
	The $SU(2)_L\times SU(2)_R$ $(\half, \half)$ representation is 
	$\Phi\equiv\frac{1}{\sqrt{2}}\left[H + i \vec{\sigma}\cdot\vec{\pi}\right]$, 
	while the $SU(2)_L\times U(1)_Y$ doublet in this paper is  
	$\phi \equiv  \Phi \left[\begin{array}{c} 1 \\0\end{array}\right]$.
	Including all 
	${\cal O}(\Lambda^2), {\cal O}(\ln \Lambda^2)$ divergences:
	\begin{eqnarray}
		\label{eqn:VGML_Ren}
		 L^{Eff;All-loops;}_{GML} &=&  \half Tr\vert\partial_\mu\Phi\vert^2-V_{GML}^{Eff} \nonumber \\
		 V_{GML}^{Eff;All-loops} &=& \frac{\lambda_\phi^2}{4}\left[ H^2+{\vec \pi}^2  - 
		 	\left(\HVEV^2 - { \frac{\mpisq}{\lambda_\phi^2} } \right) \right]^2 \nonumber \\
		&-& \HVEV\mpisq H +  {\cal O}_{Ignore}^{GML}  \,.
	\end{eqnarray}
	causes tadpoles to vanish identically, so that separate tadpole renormalization is un-necessary.
	${\cal O}_{Ignore}^{GML}$ denotes finite operators that do not contribute to UVQD,
	\begin{eqnarray}
	\label{GMLIgnore}
	\quad \quad {\cal O}_{Ignore}^{GML}
	={\cal O}^{GML}_{D>4}+{\cal O}^{GML}_{D\leq 4;NonAnalytic}+{\cal O}^{GML}_{1/ \Lambda^2;Irrelevant}\,. \nonumber
	\end{eqnarray}
	The effective potential (\ref{eqn:VGML_Ren}) 
	reduces to the 3 effective potentials of the Schwinger model \cite{Schwinger1957} as: 
	$\HVEV \to 0,\mpisq \neq 0$ (Schwinger Wigner mode); 
	$\HVEV \to 0,\mpisq \to 0$ (Schwinger scale-invariant point); 
	or $\HVEV \neq 0,\mpisq \to 0$ (Schwinger Goldstone mode).
	Ref. \cite{Lynnetal2012} extended (\ref{eqn:VGML_Ren}) to include SM quarks and leptons, 
	but possible IR divergences, 
	due to massless SM neutrinos, 
	were out of scope and ignored. 
}
, one of those T-Matrix WTI is equivalent to the Goldstone theorem in this global theory.

We use ``pion-pole dominance" arguments 
to derive these axial-vector WTIs for the SSB $\GlnuDSM$ 
in Appendix \ref{WTIproof}, 
and so rely on the masslessness of the NGB in Goldstone mode.
In this global theory 
(although not in the gauge theories that we consider in \cite{LSS-3Proof,SU(2)Proof,LSS-4Proof}) 
that masslessness translates precisely 
into the masslessness of the pseudoscalar boson $\mpisq=0$ when $\HVEV\neq0$.
$L_{\nu_dSM_{tb\tau\nu_\tau}^G}\vert_{\HVEV\neq0}$ is the subject of the remainder of this Section.

These global axial-vector WTI for the $\GlnuDSM$ are a generalization
of the classic work of B.W. Lee \cite{Lee1970},  
who constructed the all-loop-orders renormalized tower of quantum WTIs 
for the $SU(2)_L\times SU(2)_R$ Gell-Mann Levy (GML) model \cite{GellMannLevy1960} 
with Partially Conserved Axial-vector Currents (PCAC).
We replace GML's strongly-interacting \LSM~ with a weakly-interacting BEH \LSM: 
$\sigma \to H, {\vec \pi}\to {\vec \pi}, m_{\sigma}\to m_{h}, f_{\pi} \to \HVEV$;
we eliminate the explicit symmetry breaking of PCAC ($\gamma= 0$),
and reduce the symmetry from $SU(2)_L\times SU(2)_R$ to $SU(2)_L\times U(1)_Y$
when we add SM fermions and their attendant Yukawa couplings.
We also introduce a quark $SU(3)_C$, so that the resultant generation has SM couplings, which ensures that our WTI have zero axial anomaly.

\subsection{Axial-vector Ward-Takahashi identities in $\GlnuDSM$ }
\label{GlnuDSMWTIs}
We focus on the global isospin axial-vector  current ${\vec J}^{\mu}_{L-R;\GlnuDSM}$. 
The global color $SU(3)_C$, $SU(2)_{L+R}$ and electromagnetic $J_{QED}^\mu$ currents are vector currents
and are not spontaneously broken, 
so they do not yield further WTI information of interest to this paper.
{In Appendix \ref{WTIproof}, we describe how CP conservation 
enables us to consider amplitudes of the axial-vector  current, and derive towers of WTI.
In future work, we will consider the generalization to the case where CP is violated.}

Because we are interested in global-symmetric relations 
among 1-scalar-particle-irreducible (1-$\phi$-I) connected amputated Green's functions (GF) 
with external $\phi$ scalars, 
it is convenient to use tools (eg. canonical quantization) 
from Vintage Quantum Field Theory (V-QFT),
a name coined by Ergin Sezgin.  
Analysis is done in terms of the exact renormalized interacting $\GlnuDSM$ fields, 
which asymptotically become the in/out states, 
i.e. free fields for physical S-Matrix elements.
Appendix \ref{WTIproof} gives details of the derivation 
of our rigid axial-vector WTIs,
some highlights of which we present in this section.
	
For $\HVEV \neq 0$, the pseudoscalars ${\vec \pi}$ are massless%
\footnote{\label{foot:Unitary}
	The masslessness of $\vec \pi$, $\mpisq=0$, in Goldstone Mode, 
	is closely related to the masslessness 
	of the Nambu-Goldstone bosons of the broken global symmetry in this ungauged theory.
	To identify the NGBs, we must pass from the linear representation (\ref{SMHiggs})
	to  the unitary Kibble representation \cite{Ramond2004,Georgi2009},
	with transformed fields $\tilde H$ and $\vec{\tilde \pi}$, 
	and VEVs $\langle \tilde H\rangle =\HVEV$ and $\langle {\vec{\tilde\pi}} \rangle =0$,
		\begin{eqnarray}
		\label{UnitaryHiggs}
		\phi&=& \frac{1}{\sqrt{2}} {\tilde H} U;  \quad U \equiv \exp \left[ i \frac {{\vec \sigma} \cdot \vec{\tilde \pi}}{\HVEV} \right]\left[\begin{array}{c} 1 \\0\end{array}\right] \,.
		\end{eqnarray}
	$\vec {\tilde \pi}$ (not $\vec\pi$) are the purely derivatively coupled NGBs.
	We note that it is the ability to transform to the unitary representation that makes
	$\vec {\tilde \pi}$ derivatively coupled. 
	The transformation (\ref{UnitaryHiggs}) 
	is not possible in Wigner mode nor at the scale-invariant point.  
	In the ungauged theory 
	it is also possible to add a Polkinghorne PCAC term $\gamma H$ 
	that explicitly violates the axial symmetry. 
	In \cite{Lynnetal2012} we follow \cite{Lee1970} 
	in considering the $\vert\HVEV\vert$ vs. $\mpisq$ quarter-plane, 
	in which the Wigner mode is the x-axis ($\HVEV=0$),
	the Goldstone mode is the y-axis ($\mpisq=0$),
	the scale-invariant point is the origin,
	and the symmetry is explicitly broken off these axes.
	Lee \cite{Lee1970} points out a remarkable WTI: $\gamma=\HVEV\mpisq$.
	The Goldstone mode and the Wigner mode are thus just the $\mpisq\to0$
	and $\HVEV\to0$ limits of the explicitly broken theory.
	In the quarter plane, 
	the transformation to the Kibble representation 
	finds that $\vec {\tilde \pi}$ has non-derivative couplings, 
	including a mass, proportional to $\mpisq$.
	These all vanish in the $\mpisq\to0$ limit, 
	and the Goldstone theorem (among other WTIs) is recovered.
	This connection between $\mpisq=0$ and $m_{NGB}^2=0$
	appears to be severed in the gauge theory where the explicit breaking term
	is thought to be forbidden by unitarity.
}. 
We therefore solve/obey the axial-vector ``pion-pole-dominance"  
Transition-Matrix 
(T-matrix, which recall is related to the better-known S matrix by $S=1+iT$)
identity proved in Appendix \ref{WTIproof}.
\begin{eqnarray}
\label{TMatrixID}
&&\HVEV T_2^{t_1...t_Mt}(p_1 ...p_N;q_1...q_M0) \nonumber  \nonumber \\
&&\quad \quad =\sum ^M_{m=1} \delta^{tt_m} T^{t_1...{\widehat {t_m}}...t_M}(p_1...p_Nq_m;q_1...{\widehat {q_m}}...q_M) \nonumber \\
&&\quad \quad -\sum ^N_{n=1} T^{t_1...t_Mt}(p_1 ...{\widehat {p_n}}...p_N;q_1...q_Mp_n) 
\end{eqnarray}
with N renormalized $h=H-\HVEV$ external legs (coordinates x, momenta p), 
and M renormalized ($CP=-1$) ${\vec \pi}$ external legs (coordinates y, momenta q, isospin t). 

{ Equation (\ref{TMatrixID}) 
either relates T-matrix elements all with even numbers of $\vec \pi$ (if $M$ is odd),
or T-matrix elements all with odd numbers of $\vec \pi$ (if $M$ is even).
 Because $CP$ is conserved, T-matrix elements  with odd numbers of $\vec \pi$ vanish,
hence (\ref{TMatrixID}) is of interest only for $M$ odd.}

Here $T\equiv T_1+T_2$. 
$T_1$ includes only diagrams 
with an extra zero-momentum external leg ${\vec \pi}$, 
attached directly to an external $h$ or another external $\vec \pi$ leg 
as in Figure 1.
The notation  ${\widehat {p_n}},{\widehat {q_m}},{\widehat {t_m}}$ indicates that ``hatted'' external fields and momenta are to be omitted.

\begin{figure}
\centering
\includegraphics[width=1\hsize,trim={0cm 5cm 0cm 5cm}]{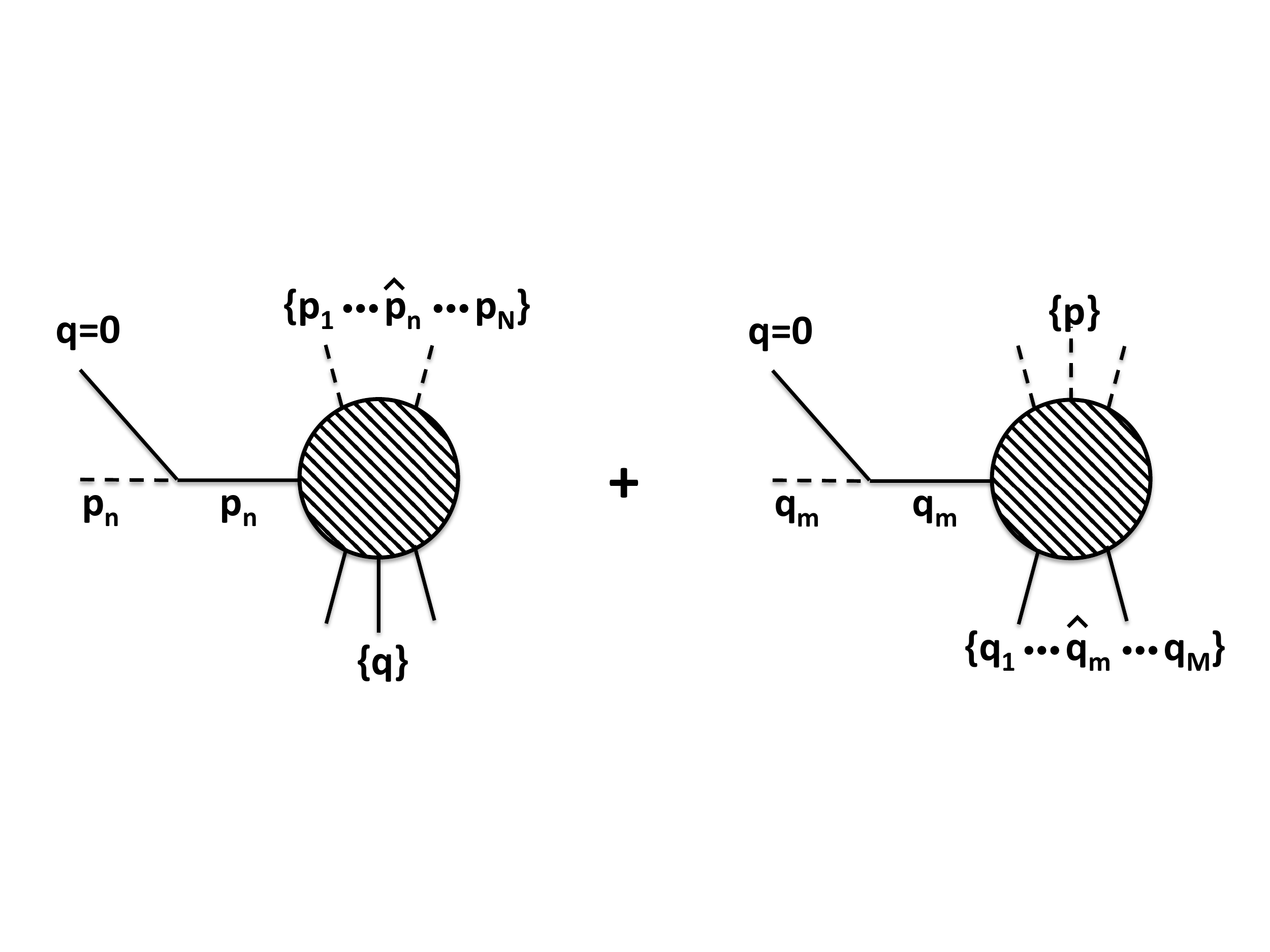}
\caption{
\label{fig:LeeFig10} 
$ T_1^{t_1...t_Mt}(p_1 ...p_N;q_1...q_M0) $:
Hashed circles are 1-$\phi$-R $T^{t_1...t_M}(p_1 ...p_N;q_1...q_M) $, solid lines $\vec \pi$, dashed lines $h$.
One (zero-momentum) soft pion is attached to an external leg (i.e. a branch)  in all possible ways.
Fig. \ref{fig:LeeFig10}  is the $SU(3)_C\times SU(2)_L\times U(1)_Y$  $\GlnuDSM$  analogy of B.W. Lee's  Figure 10 \cite{Lee1970}.
}
\end{figure}

A tower of quantum  WTI recursion relations, 
among renormalized connected amputated 1-Scalar-Particle-Irreducible (1-$\phi$-I) Greens Functions (GF) 
$\Gamma_{N,M}^{t_1...t_M}(p_1...p_N;q_1...q_M)$,
with N external renormalized $h=H-\HVEV$ (coordinates x, momenta p), 
and M external ($CP=-1$) renormalized ${\vec \pi}$ (coordinates y, momenta q, isospin t), 
is shown in Appendix \ref{WTIproof} to be a solution to the T-Matrix identity (\ref{TMatrixID}). 
The resulting WTI
relate a 1-$\phi$-I connected amputated GF with $(N+M+1)$ external fields, 
including an extra zero-momentum ${\vec \pi}$, 
to two 1-$\phi$-I amputated GFs with $(N+M)$ external fields. 
For $\vec \pi$ with $CP=-1$, the result
\begin{eqnarray}
	\label{eqn:BWLeeWTI}
	&&\HVEV\Gamma_{N,M+1}^{t_1...t_Mt}(p_1 ...p_N;q_1...q_M0) \nonumber  \nonumber \\
	&&\quad \quad =\sum ^M_{m=1} \delta^{tt_m}\Gamma_{N+1,M-1}^{t_1...{\widehat {t_m}}...t_M}
(p_1...p_Nq_m;q_1...{\widehat {q_m}}...q_M) \nonumber \\
&&\quad \quad -\sum ^N_{n=1}\Gamma_{N-1,M+1}^{t_1...t_Mt}(p_1 ...{\widehat {p_n}}...p_N;q_1...q_Mp_n)
\end{eqnarray} 
is valid for $N,M \ge0$, { though non-trivial only for odd $M$.} 
(Hatted quantities are again omitted.)

We form the $\phi$-sector effective Lagrangian as a sum
\begin{equation}
\label{eqn:CompleteEffectiveLagrangian}
L^{Eff;\GlnuDSM}_{\phi} = \sum_{N,M}L^{Eff;\GlnuDSM}_{\phi;N,M}
\end{equation}
over all possible numbers of external scalars $h$ and pseudoscalars $\pi^i$.
Each term, $L^{Eff;\GlnuDSM}_{\phi;N,M}$ is obtained by attaching to $\Gamma_{N,M}^{t_1...t_M}$: 
$N$ appropriate external scalar wavefunctions; 
$M$ appropriate external pseudoscalar wavefunctions, with sums over isospins; 
and combinatoric factors for identical external  boson fields $h,{\vec \pi}$.

{\em It is worth emphasing that 
all perturbative quantum loop corrections, to all-loop-orders, 
are included in the $\phi$-sector effective Lagrangian:}
1-$\phi$-I connected amputated GF $\Gamma_{N,M}^{t_1...t_M}(p_1...p_N;q_1...q_M)$ in (\ref{eqn:BWLeeWTI}); wavefunction renormalizations;  
renormalized scalar propagators (\ref{pNGBPropagator},\ref{BEHPropagator}); 
the BEH VEV $\HVEV$.
Eqn. (\ref{eqn:BWLeeWTI}) includes the full set of quantum all-loop-orders 
from the global $SU(3)_{C}\times SU(2)_L \times U(1)_Y$ theory,
originating in loops containing virtual $\GlnuDSM$: 
quarks $q_L^{c},t_R^{c};b_R^{c}$ and leptons $l_L,{\nu_\tau}_R;\tau_R$,
with colors $c=r,w,b$; 
and scalars $h,{\vec \pi}$. 
Because they arise entirely from global axial-vector WTI, 
our results are independent of regularization-scheme \cite{KrausSiboldAHM}.

There remains, however, one more crucial step. 
{\it We must (!) impose all those symmetries of the 1-$\phi$-R T-matrix 
that are not symmetries of the 1-$\phi$-I Green's functions (\ref{eqn:BWLeeWTI}) 
nor of the complete effective Lagrangian 
(\ref{eqn:CompleteEffectiveLagrangian}).}
\footnote{
	Failing to impose those T-matrix symmetries 
	(eg. crucially the one that is equivalent to the Goldstone Theorem) 
	results in a {\bf mistake}. 
}
In particular the $\GlnuDSM$-analogue of the Adler self-consistency conditions
\cite{Adler1965,AdlerDashen1968}  
(see for example \cite{Lee1970} p. 37), derived in Appendix A, 
of which the Goldstone Theorem is a special case,
ensures the infra-red finiteness of the theory 
for exactly zero pseudoscalar masses, $\mpisq=0$.

\subsection{Construction of SM scalar-sector effective Lagrangian from axial-vector Ward-Takahashi IDs}
\label{EffectiveWTILagrangianSM}
We want to classify operators arising from $\GlnuDSM$ degrees of freedom,
and separate the finite operators from the divergent ones. 
There are finite operators that arise entirely from $\GlnuDSM$ degrees of freedom.   
Although important for computing ``physical observables" in the $\GlnuDSM$ 
	\big(e.g. the analogy of the successful 1-loop high-precision Standard Model  
	predictions for the top-quark from
	Z-pole physics \cite{LynnStuart1985,Nobel1999} in 1984
	and the $W^{\pm}$ mass \cite{Sirlin1980,Nobel1999} in 1980, 
as well as the 2-loop BEH mass  from
	Z-pole physics \cite{LynnStuart1985,Verzegnassi1987,Nobel2013}
	and the $W^{\pm}$ mass \cite{Sirlin1980,Verzegnassi1987,Nobel2013}\big),
they are not the point of this paper. 
We want instead to focus on UVQD, logarithmic UV divergences, 
and finite relevant operators, to see how they are related by the WTIs.
The reader might imagine ${\cal O}(\Lambda^2)$ and ${\cal O}(\ln \Lambda^2)$ divergences, 
never taking the limit $\Lambda^2 \to \infty$.  

In actuality, 
there are 3 classes of {\it finite} operators in the $\GlnuDSM$
that we will ignore: 
\begin{itemize}
\item finite  ${\cal O}^{\GlnuDSM}_{1/\Lambda^2;Irrelevant}$ vanish as $m_{Weak}^2/ \Lambda^2 \to 0$;
\item ${\cal O}^{\GlnuDSM}_{Dim>4}$ are finite dimension $>4$ operators;  
\item ${\cal O}^{\GlnuDSM}_{Dim\leq4;NonAnalytic}$ are finite dimension$\leq4$ operators 
that are non-analytic in momenta or in a renormalization scale $\mu^2$. 
\end{itemize}
Such finite operators appear throughout the axial-vector Ward-Takahashi IDs (\ref{eqn:BWLeeWTI}): 
\begin{itemize}
\item $N+M \geq 5$ gives relations among  ${\cal O}^{\GlnuDSM}_{1/ \Lambda^2; Irrelevant}$ 
and ${\cal O}^{\GlnuDSM}_{Dim>4}$;
\item the left hand side of  (\ref{eqn:BWLeeWTI}) for $N+M=4$ 
is also ${\cal O}^{\GlnuDSM}_{Dim>4}$ or ${\cal O}^{\GlnuDSM}_{1/ \Lambda^2; Irrelevant}$;
\item  $N+M\leq 4$ operators ${\cal O}^{\GlnuDSM}_{Dim\leq 4;NonAnalytic}$ 
also appear in those WTI. 
\end{itemize}
All such operators  will be ignored below:
\begin{eqnarray}
{\cal O}^{\GlnuDSM}_{Ignore}&=&{\cal O}^{\GlnuDSM}_{1/\Lambda^2;Irrelevant} \nonumber
\\&+&{\cal O}^{\GlnuDSM}_{Dim>4}+{\cal O}^{\GlnuDSM}_{Dim\leq4;NonAnalytic}
\end{eqnarray}

Finally, there are  $N+M\leq 4$ operators that are analytic in momenta. 
We expand these in powers of momenta, 
count the resulting dimension of each term in the operator Taylor-series, 
and ignore ${\cal O}^{\GlnuDSM}_{Dim>4}$ 
and ${\cal O}^{\GlnuDSM}_{1/ \Lambda^2; Irrelevant}$ terms.

We seek next to classify the relevant operators, 
in this case the ${\vec \pi}$ and $h$ inverse propagators (together with  tadpoles).

Define the exact renormalized pseudoscalar propagator (no sum on $j$) 
in terms of a $\vec \pi$ pole, 
the K$\ddot a$ll$\acute e$n-Lehmann spectral density $\rho_{\pi}$ \cite{Bjorken1965,ItzyksonZuber}, 
and  wavefunction renormalization. We assume ${\vec \pi}$ decays weakly: 
\begin{eqnarray}
\label{pNGBPropagator}
&&\Delta_{\pi}(q^2) = 
    -i(2\pi)^2\langle 0\vert T\left[ \pi_j(y)\pi_j(0)\right]\vert 0
          \rangle\vert^{Fourier}_{Transform} \nonumber \\
&&\quad \quad = 
     \frac{1}{q^2-m_{\pi;Pole}^2 + i\epsilon} 
     + \int dm^2 \frac{\rho_{\pi}(m^2)}{q^2-m^2 + i\epsilon} \nonumber \\
&&Z_{\phi}^{-1} = 1+ \int dm^2 \rho_{\pi}(m^2) 
\end{eqnarray} 

Define similarly the BEH scalar propagator 
in terms of a BEH scalar pole, the spectral density $\rho_{BEH}$, 
and the {\it same} wavefunction renormalization. 
We assume $h$ also decays weakly and resembles a resonance:
\begin{eqnarray}
\label{BEHPropagator}
&&\Delta_{BEH}(q^2) = 
	-i (2\pi)^2\langle 0\vert T\left[ h(x) h(0)\right]\vert 0
	\rangle\vert^{Fourier}_{Transform} \nonumber \\
&& \quad \quad =\frac{1}{q^2-m_{h;Pole}^2 + i\epsilon}
	+ \int dm^2 \frac{\rho_{BEH}(m^2)}{q^2-m^2 + i\epsilon} \nonumber \\
&& Z_{\phi}^{-1} = 1+ \int dm^2 \rho_{BEH}(m^2)  \nonumber \\
&& \int dm^2 \rho_{\pi}(m^2) = \int dm^2 \rho_{BEH}(m^2) 
\end{eqnarray}
The connected amputated 1-$\phi$-I ${\vec \pi}$ and $h$ inverse propagators are:
\begin{eqnarray}
\label{InversePropagators}
\Gamma_{0,2}^{t_1t_2}(;q,-q) &\equiv& \delta^{t_1t_2} \Gamma_{0,2}(;q,-q) \nonumber \\
\Gamma_{0,2}(;q,-q) &\equiv& \left[ \Delta_{\pi}(q^2) \right]^{-1}  \\
\Gamma_{2,0}(q,-q;) &\equiv& \left[ \Delta_{BEH}(q^2) \right]^{-1} \nonumber
\end{eqnarray}

The spectral density parts of the propagators 
\begin{eqnarray}
\label{SpectralDensityPropagators}
 \Delta_{SpectralDensity}^{BEH}(q^2) &=& \int dm^2 \frac{\rho_{BEH}(m^2)}{q^2-m^2 + i\epsilon} \\
 \Delta_{SpectralDensity}^{\vec \pi}(q^2) &=& \int dm^2 \frac{\rho_{\pi}(m^2)}{q^2-m^2 + i\epsilon}  \nonumber
\end{eqnarray}
are clearly finite. 
From dimensional analysis of (\ref{pNGBPropagator},\ref{BEHPropagator}), 
the contribution of a state  of mass/energy $\sim M_{Heavy}$ 
to the spectral densities $\rho_{\pi}(M_{Heavy}^2)$ and $\rho_{BEH}(M_{Heavy}^2)$, 
and  to $\Delta_{SpectralDensity}^{BEH},\Delta_{SpectralDensity}^{\vec \pi}$, 
scale as $M_{Heavy}^{-2}$.
The Euclidean cut-off therefore contributes only $\sim \frac{1}{\Lambda^2}$.

We now form the
all-loop-orders renormalized scalar-sector effective Lagrangian for ($h,\vec \pi$) with CP=($1,-1$) 
\begin{eqnarray}
\label{FormSchwingerPotential}
&& L^{Eff}_{\phi;\nu_D SM^G_{tb\tau\nu_\tau}} =  \Gamma_{1,0}(0;)h +\frac{1}{2!} \Gamma_{2,0}(p,-p;)h^2 \nonumber \\
&&  \quad \quad + \frac{1}{2!} \Gamma_{0,2}^{t_1t_2}(;q,-q)\pi_{t_1}\pi_{t_2} +\frac{1}{3!} \Gamma_{3,0}(000;)h^3  \nonumber \\ 
&& \quad \quad + \frac{1}{2!} \Gamma_{1,2}^{t_1t_2}(0;00) h \pi_{t_1}\pi_{t_2} +\frac{1}{4!} \Gamma_{4,0}(0000;)h^4  \nonumber \\
&&  \quad \quad + \frac{1}{2!2!} \Gamma_{2,2}^{t_1t_2}(00;00) h^2 \pi_{t_1}\pi_{t_2}  \\
&& \quad \quad +  \frac{1}{4!}\Gamma_{0,4}^{t_1t_2t_3t_4}(;0000)\pi_{t_1}\pi_{t_2} \pi_{t_3}\pi_{t_4} + {\cal O}_{\nu_D SM^G_{tb\tau\nu_\tau}}^{Ignore}  \nonumber 
\end{eqnarray} 

The connected amputated Green's function identities (\ref{eqn:BWLeeWTI})
severely constrain the effective Lagrangian (\ref{FormSchwingerPotential}). 
For pedagogical clarity, we first separate out the isospin indices
\begin{eqnarray}
\label{IsospinIndices}
	\Gamma_{0,2}^{t_1t_2}(;q,-q) &\equiv& \delta^{t_1t_2} \Gamma_{0,2}(;q,-q)\,, \nonumber \\
	\Gamma_{1,2}^{t_1t_2}(-q;q0) &\equiv& \delta^{t_1t_2}\Gamma_{1,2}(-q;q0) \,,\nonumber  \\
	\Gamma^{t_1t_2}_{2,2}(00;00)&\equiv&\delta^{t_1t_2}\Gamma_{2,2}(00;00)\,, \\
	 \Gamma_{0,4}^{t_1t_2t_3t_4}(;0000)  &\equiv& \Gamma_{0,4}(;0000) \nonumber \\
	&\times& \left[ \delta^{t_1t_2}\delta^{t_3t_4} + \delta^{t_1t_3}\delta^{t_2t_4} + \delta^{t_1t_4}\delta^{t_2t_3} \right]\,. \nonumber 
\end{eqnarray}

Itemizing the relevant WTI and their effects on (\ref{FormSchwingerPotential}), 
setting momenta to zero except where needed, suppressing the isospin indices, and indicating
the finite operators as simply ${\cal O}_{Ignore}$:
\begin{itemize}
\item WTI $N=0, M=1$
\begin{eqnarray}
\label{NM01}
\delta^{t_1t_2}\Gamma_{1,0}(q;)  &=& \HVEV \Gamma_{0,2}^{t_1t_2}(;q,-q)\,,\nonumber \\
\Gamma_{1,0}(0;) &=& \HVEV \Gamma_{0,2}(;00) \,,
\end{eqnarray}
since no momentum can run into the tadpoles.
\item WTI $N=1, M=1$
\begin{eqnarray}
\label{NM11}
\delta^{t_1t_2}\Gamma_{2,0}(-q,q;) &-& \Gamma_{0,2}^{t_1t_2}(;q,-q)  \nonumber \\
&=&\HVEV \Gamma_{1,2}^{t_1t_2}(-q;q0) \,, \nonumber \\
\Gamma_{2,0}(-q,q;) &-& \Gamma_{0,2}(;q,-q)   \\
&=&\HVEV \Gamma_{1,2}(-q;q0)  \nonumber \\
&=&\HVEV \Gamma_{1,2}(0;00)  + {\cal O}^{\nu_D SM^G}_{Ignore} \nonumber \,,\\ 
\Gamma_{2,0}(00;) &=& \Gamma_{0,2}(;00) + \HVEV \Gamma_{1,2}(0;00)\nonumber 
\end{eqnarray}
\item WTI $N=2, M=1$
\begin{eqnarray}
\label{NM21}
\HVEV\Gamma^{t_1t_2}_{2,2}(00;00) &=& \delta^{t_1t_2}\Gamma_{3,0}(000;) -2\Gamma^{t_1t_2}_{1,2}(0;00) \,, \nonumber \\
\HVEV\Gamma_{2,2}(00;00) &=& \Gamma_{3,0}(000;) -2\Gamma_{1,2}(0;00) \,.
\end{eqnarray}
\item WTI $N=0, M=3$
\begin{eqnarray}
\label{NM03}
-\HVEV\Gamma^{t_1t_2t_3t_4}_{0,4}(;0000) &=& \delta^{t_1t_2}\Gamma_{1,2}^{t_3t_4}(0;00)   \\
&+& \delta^{t_1t_3}\Gamma_{1,2}^{t_2t_4}(0;00)+ \delta^{t_1t_4}\Gamma_{1,2}^{t_2t_3}(0;00)\,, \nonumber \\
-\HVEV\Gamma_{0,4}(;0000) &=& \Gamma_{1,2}(0;00)\,.\nonumber
\end{eqnarray}
\item WTI $N=1, M=3$
\begin{eqnarray}
\label{NM13}
\delta^{t_1t_2}\Gamma^{t_3t_4}_{2,2}(00;00) &+& \delta^{t_1t_3}\Gamma^{t_2t_4}_{2,2}(00;00) +\delta^{t_1t_4}\Gamma^{t_2t_3}_{2,2}(00;00) \nonumber \\
&-& \Gamma^{t_1t_2t_3t_4}_{0,4}(;0000) =0 \,, \\
\Gamma_{2,2}(00;00)&=&\Gamma_{0,4}(;0000) \,. \nonumber
\end{eqnarray}
\item WTI $N=3, M=1$
\begin{eqnarray}
\label{NM31}
-\delta^{t_1t_2}\Gamma_{4,0}(0000;)&+&3\Gamma^{t_1t_2}_{2,2}(00;00)=0 \,, \\
-\Gamma_{4,0}(0000;)&+&3\Gamma_{2,2}(00;00)=0 \,.\nonumber
\end{eqnarray}
\end{itemize}

The quadratic and quartic coupling constants are defined in terms of 2-point and  4-point 1-$\phi$-I connected amputated GF
\begin{eqnarray}
\label{SchwingerWignerData}
  \Gamma_{0,2}(;00)  &\equiv& -\mpisq \\
\Gamma_{0,4}(;0000)  &\equiv& -2 \lambda_{\phi}^2 \,. \nonumber 
\end{eqnarray} 

The pseudoscalar and $h$ (BEH) scalar masses are most usefully defined as: 
\begin{eqnarray}
	\label{MassRelationpi}
	\mpisq &\equiv& -\Gamma_{0,2}(;00)  =  \Bigg[ \frac{1}{m_{\pi ;Pole}^2}+ \int dm^2 \frac{\rho_{\pi}(m^2)}{m^2 } \Bigg]^{-1} \\
	m_{h}^2 &\equiv&  -\Gamma_{2,0}(00;)  =  \Bigg[ \frac{1}{m_{h ;Pole}^2}+ \int dm^2 \frac{\rho_{BEH}(m^2)}{m^2 } \Bigg]^{-1}\,. \nonumber 
\end{eqnarray}

The third $N=1$, $M=1$ WTI of equation (\ref{NM11})
can then be rewritten instructively as 
a mass-relation between the BEH $h$ scalar 
and the 3 pseudoscalar bosons ${\vec \pi}$: 
\begin{eqnarray}
\label{MassRelationBEH}
m_h^2 &=& \mpisq + 2\lambda_{\phi}^2 \HVEV ^2 \,,
\end{eqnarray}
a more familiar form
which we have employed in previous papers \cite{Lynn2011,Lynnetal2012,LynnStarkman2013}.

The  all-loop-orders renormalized $\phi$-sector effective Lagrangian (\ref{FormSchwingerPotential}), 
 constrained {\bf only} by those axial-vector WTIs governing Greens functions (\ref{eqn:BWLeeWTI}),  
may be written
\begin{eqnarray}
\label{LEffectiveSM}
L^{Wigner;SI;Goldstone} &=& L^{Kinetic} 
	- V^{Wigner;SI;Goldstone}\nonumber \\
&+& {\cal O}_{Ignore}^{\GlnuDSM} \,.
\end{eqnarray}
The kinetic term incorporates the  non-trivial (but finite) wavefunction renormalization
\begin{eqnarray}
L^{Kinetic}&=&\half \Big( \Gamma_{0,2}(;p,-p) - \Gamma_{0,2}(;00) \Big) h^2  \\
&+&\half \Big( \Gamma_{0,2}(;q,-q)- \Gamma_{0,2}(;00)  \Big) {\vec \pi}^2\,,\nonumber 
\end{eqnarray}
with
\begin{equation}
\Gamma_{0,2}(;q,-q)-\Gamma_{0,2}(;00) \sim q^2\,,
\end{equation}
while the effective potential
\begin{eqnarray}
\label{WignerSIGoldstonePotential}
V^{Wigner;SI;Goldstone} &=& 
\mpisq \Big[ \frac{h^2 + {\vec \pi}^2}{2} +\HVEV h \Big]  \\
&+& \lambda_{\phi}^2 \Big[ \frac{h^2 + {\vec \pi}^2}{2} +\HVEV h \Big]^2 \nonumber
\end{eqnarray} 
incorporates all 3 modes (i.e. Wigner mode, Scale Invariant and Goldstone mode) 
of the Lagrangian (\ref{SMFieldsLagrangian}).\footnote{
\label{InstructiveFootnote}
	It is instructive, and we argue \cite{Lynn2011Updated} dangerous, to 
	ignore vacuum energy and re-write the potential in (\ref{WignerSIGoldstonePotential}) as:
	\begin{eqnarray}
	\label{VViolatesGoldstoneTheoremExtendedAHM}
	V^{Wigner;SI;Goldstone}_{\nu_D SM^G} &=& {\lambda_\phi^2}\left[ \phi^\dagger \phi - \half \left(\HVEV^2 - { \frac{\mpisq}{\lambda_\phi^2} } \right) \right]^2 \quad \quad
	\end{eqnarray} 
	using $\frac{h^2 + {\vec \pi}^2}{2} +\HVEV h = \phi^\dagger \phi - \half \HVEV^2$.
	If one then minimizes 
	$V^{Wigner;SI;Goldstone}_{\phi} $ 
	while ignoring the crucial constraint imposed by the Goldstone Theorem,
	(or more precisely by the WTI 
	that is equivalent to the Goldstone Theorem in this ungauged theory: 
	 see Subsection \ref{IRFiniteness}),
	the resultant (incorrect and un-physical) minimum
	$ \big< H \big>_{FT}^2 \equiv \Big( \HVEV^2 - \frac{\mpisq} {\lambda_\phi^2}  \Big)  $ 
	does not distinguish properly between the 3 modes (\ref{WignerSIGoldstonePotentialsExtendedAHM}) of 
	(\ref{VViolatesGoldstoneTheoremExtendedAHM}).
	At issue is the 
	renormalized
	\begin{eqnarray}
	\label{FTExtendeAHM}
	\mpisq &=& \mu_{\phi;Bare}^2 +C_\Lambda \Lambda^2+C_{BEH} m_{BEH}^2 +\delta \mpisq \nonumber \\
	&+&C_{Heavy} M_{Heavy}^2+C_{Heavy;ln}M_{Heavy}^2 \ln{(M_{Heavy}^2)}  \nonumber \\
	&+&C_{Heavy;\Lambda} M_{Heavy}^2\ln{(\Lambda^2)} +\lambda_\phi^2 \HVEV^2
	\end{eqnarray}
	where the $C$'s are constants, and $m_h^2=\mpisq +2\lambda^2\HVEV^2$. 
	It is ``fashionable" to simply drop the UVQD term $C_\Lambda \Lambda^2$ in (\ref{FTExtendeAHM}), and argue 
	that it is somehow an artifact of dimensional regularization (DR), 
	even though M.J. G. Veltman \cite{Veltman1981} showed 
	that UVQD {\em do} appear at 1-loop in the SM, 
	and are properly handled by DR's poles at dimension $Dim=2$.  
	We keep UVQD. 
	\newline \indent
	For pedagogical efficiency, 
	we have included in  (\ref{FTExtendeAHM}) 
	terms with $M_{Heavy}^2 \gg m_{Weak}^2$, 
	such as might arise in BSM physics
	(cf. Section \ref{RenormalizationBSM}).
	\newline \indent
	In Wigner mode, where $\HVEV =0$:
	\begin{eqnarray}
	\label{WignerDimRegFTExtendedAHM}
	m_{h}^2 &=& m_{\pi}^2 \sim \Lambda^2,M_{Heavy}^2 \gg m_{Weak}^2 \,.
	\end{eqnarray} 
	During renormalization of a tree-level weak-scale BEH mass-squared 
	$m_{h;Bare}^2 \sim m_{Weak}^2$, 
	relevant operators originating in quantum loops 
	appear to ``naturally" force the renormalized value 
	up to the heavy scale (\ref{WignerDimRegFTExtendedAHM}). 
	Wigner mode is therefore quantum-loop unstable, 
	because the heavy scale cannot decouple from the weak scale!
	Eqn. (\ref{WignerDimRegFTExtendedAHM}) is 
	the motivation for much BSM physics, 
	even though our universe is not in Wigner mode.
	\newline \indent
	In the {Spontaneously broken Goldstone  mode}, 
	where $\HVEV \neq0$, 
	in obedience to a WTI (equivalent to the Goldstone Theorem) 
	in Subsection \ref{IRFiniteness} below, 
	the bare counter-term $\mu_{\phi;Bare}^{2}$ in (\ref{FTExtendeAHM}) is defined by 
	\begin{eqnarray}
	\label{GoldstoneFTExtendedAHM}
	\mpisq  \equiv 0 \,.
	\end{eqnarray} 
	We show in Subsection \ref{IRFiniteness} that, 
	for constant $\vec \theta$, 
	the zero-value in  (\ref{GoldstoneFTExtendedAHM}) 
	is protected by the NGB shift symmetry 
	\begin{eqnarray}
	\label{BEHFTShiftSymmetry}
	{\tilde {\vec \pi}}\to {\tilde {\vec \pi}}+{\tilde {\vec \pi}}\times {\vec \theta} + \HVEV {\vec \theta} + {\cal O}(\theta^2) \,.
	\end{eqnarray}
	\newline\indent
	Minimization of (\ref{VViolatesGoldstoneTheoremExtendedAHM})
	violates stationarity of the true minimum at $\HVEV$\cite{ItzyksonZuber} 
	and destroys the theory's renormalizability and unitarity, 
	which require that dimensionless wavefunction  renormalization 
	$\HVEV_{Bare}=\Big[ Z^{\phi}\Big] ^{1/2}\HVEV$ 
	contain no relevant operators \cite{Lynn2011,Bjorken1965,ItzyksonZuber}. 
	The crucial observation is that, in obedience to the Goldstone theorem, 
	$Renormalized (\HVEV_{Bare}^2) \neq  \HVEV_{FT}^2$.
	} 
The effective potential  in (\ref{WignerSIGoldstonePotential}) becomes in various limits: 
Wigner mode $(\HVEV =0;\mpisq =m_{h}^2\neq 0)$; 
$SI$=Scale-Invariant point $(\HVEV =0;\mpisq = m_{h}^2= 0)$; 
or Goldstone mode $\HVEV \neq 0;\mpisq = 0; m_{h}^2\neq 0)$;
\begin{eqnarray}
\label{WignerSIGoldstonePotentialsExtendedAHM}
V^{Wigner}&=& \mpisq \Big[ \frac{h^2 + {\vec \pi}^2}{2} \Big] + \lambda_{\phi}^2 \Big[ \frac{h^2 + {\vec \pi}^2}{2}  \Big]^2 \nonumber \\
V^{ScaleInvariant}&=& \lambda_{\phi}^2 \Big[ \frac{h^2 + {\vec \pi}^2}{2}  \Big]^2 \\
V^{Goldstone} &=& \lambda_{\phi}^2 \Big[ \frac{h^2 + {\vec \pi}^2}{2} +\HVEV h \Big]^2 \nonumber
\end{eqnarray}

But (\ref{WignerSIGoldstonePotentialsExtendedAHM})  has exhausted the constraints, on the allowed terms in the $\phi$-sector effective  Lagrangian, 
due to those axial-vector WTIs which govern 1-$\phi$-I connected amputated Green's functions $\Gamma_{N,M}$. 
In order to distinguish among
the effective potentials in (\ref{WignerSIGoldstonePotentialsExtendedAHM}), 
we must turn to those axial-vector WTIs that govern 
1-$\phi$-R connected amputated  T-Matrix elements.

\subsection{Infra-red finiteness, Goldstone theorem, and automatic tadpole renormalization}
\label{IRFiniteness}
\begin{quote}
{\it ``Whether you like it or not, 
you have to include in the Lagrangian all possible terms consistent with locality and power counting, 
unless otherwise constrained by Ward identities."}
Kurt Symanzik, in a 1970 private letter to Raymond Stora \cite{SymanzikPC} 
\end{quote}

In Appendix \ref{WTIproof} we extend Adler's self-consistency condition \cite{Adler1965,AdlerDashen1968} 
(originally written for the $SU(2)_L\times SU(2)_R$ \GMLfull model \cite{GellMannLevy1960}), 
to the case of the $\GlnuDSM$ Lagrangian  (\ref{SMFieldsLagrangian})
\begin{eqnarray}
\label{Adler}
&&\lim{_{_{q_{\mu}\to 0}}}\HVEV T^{tt_1...t_M}\left(p_1...p_N;qq_1...q_M\right) \vert_{q_1^2=...q_M^2=0}^{p_1^2=...p_N^2=m^2_h} \nonumber \\
&&\qquad\equiv \HVEV T^{tt_1...t_M}\left(p_1...p_N;0q_1...q_M\right) \vert_{q_1^2=...q_M^2=0}^{p_1^2=...p_N^2=m^2_h}  \nonumber \\
&&\qquad=0
\end{eqnarray}
where, for pedagogical simplicity, we will suppress $M+1$ isospin indices in
$T^{tt_1...t_M}\left(p_1...p_N;qq_1...q_M\right)$ going forward.
The T-matrix vanishes as one of the pion momenta goes to zero, 
provided all other physical scalar particles are on mass-shell. 
These are ``1-soft-pion" theorems \cite{AdlerDashen1968}.
Eqn.
(\ref{Adler})  asserts the absence of infrared divergences 
in the physical-scalar sector in Goldstone mode 
 $\GlnuDSM$. 
``Although individual Feynman diagrams may be IR divergent, 
those IR divergent parts cancel exactly in each order of perturbation theory. 
Furthermore, the Goldstone mode amplitude must vanish in the soft-pion limit \cite{Lee1970}".

A special case of (\ref{Adler}) is  the Goldstone Theorem itself,
or at least equivalent to it --
the  $N=0,M=1$ case of (\ref{Adler}) reads
\begin{eqnarray}
\label{TMatrixGoldstoneTheorem}
\HVEV T_{0,2}\left(;00\right) =0
\end{eqnarray}
where momentum conservation forces $q_1=0$ (so that $q_1^2=0$).
We may write

\footnote{
	Recall that $T_{0,2}$ is 1-P-R, while $\Gamma_{0,2}$ is 1-P-I.  
	Consider the sum of all diagrams contributing to $T_{0,2}(;00)$ 
	to all loops.  
	Each of these diagrams has exactly two (amputated) external legs, 
	both zero-4-momentum $\pi$'s. 
	Attach to either of these external legs a $\pi$ propagator (at zero 4-momentum) $\Delta_\pi(0)$
	and a $\Gamma_{0,2}(;00)$.  
	This diagram is also a contribution to $T_{0,2}(;00)$. 
	(Indeed, we can repeat this procedure an arbitrary number of times, 
	and each resulting diagram must again be a contribution
	to $T_{0,2}(;00)$.) 
	Now $\Delta_\pi(0)$ has a pole at zero four-momentum when $m_\pi^2=0$. 
	so if these contributions to 
	$T_{0,2}(;00)$ are not to diverge, $\Gamma_{0,2}$ must vanish as the external
	momenta go to zero, so that the product $\Delta_\pi(0)\Gamma_{0,2}(;00)$ does
	not diverge.
}

(\ref{TMatrixGoldstoneTheorem})
as a further constraint on the 1-$\phi$-I connected amputated Greens functions
\begin{eqnarray}
\label{GFGoldstoneTheorem}
\HVEV \Gamma_{0,2}\left(;00\right)\equiv-\HVEV\mpisq =0
\end{eqnarray}
As described in footnote \ref{foot:Unitary} above, 
the actual Goldstone theorem states that the mass of the NGB $\tilde\pi$ vanishes,
where $\tilde\pi$ are the angular degrees of freedom in the unitary representation
of the $\Phi$ field.  
However,$m_{\tilde\pi}^2=0$ if and only if $\mpisq=0$ in this global theory.

A crucial effect of the Adler relation (\ref{TMatrixGoldstoneTheorem}),
together with the  $N=0, M=1$ Ward-Takahashi Greens function identity (\ref{NM01}), 
is to automatically eliminate tadpoles in (\ref{LEffectiveGoldstoneTheorem})
\begin{eqnarray}
\label{ZeroTadpoles}
\Gamma_{1,0}(0;) &=& \HVEV \Gamma_{0,2}(;00) =0\,,
\end{eqnarray}
so that separate tadpole renormalization is un-necessary.

With $\HVEV \neq0$, 
 (\ref{TMatrixGoldstoneTheorem}) and  (\ref{GFGoldstoneTheorem}) may be written:
\begin{eqnarray}
\label{GoldstoneTheorem}
&&-\Gamma_{0,2}(;00) \equiv  -\left[ \Delta_{\pi}(0)  \right]^{-1}  \equiv \mpisq   \\
 && \quad \quad =m_{\pi ;Pole}^2\Bigg[ 1+ m_{\pi ;Pole}^2\int dm^2 \frac{\rho_{\pi}(m^2)}{m^2 } \Bigg]^{-1} \nonumber \\
&& \quad \quad =0 \nonumber
\end{eqnarray}
The pole masses of the pseudoscalars ${\vec \pi}$ in the $\GlnuDSM$ 
therefore  vanishes exactly
\begin{eqnarray}
\label{PionPoleMass}
m_{\pi;Pole}^2 &=& \mpisq\Bigg[ 1- \mpisq\int dm^2 \frac{\rho_{\pi}(m^2)}{m^2 } \Bigg]^{-1} =0 \,.\quad \quad 
\end{eqnarray}
which is the reason that  ${\tilde {\vec\pi}}$ are Nambu-Goldstone bosons.

\subsection{$\GlnuDSM$ scalar-sector effective Lagrangian obedient to Goldstone theorem $\HVEV \neq 0,\mpisq \equiv 0$}
\label{GoldstoneSM}

We now re-write the effective Lagrangian (\ref{LEffectiveSM}) 
including the constraint from 
(\ref{GFGoldstoneTheorem}), i.e. $\mpisq=0$:
\begin{eqnarray}
\label{LEffectiveGoldstoneTheorem}
L^{Eff}_{\GlnuDSM} &=& L^{Kinetic}_{\GlnuDSM}- V^{Eff;Goldstone}_{\GlnuDSM} \nonumber \\
&+&{\cal O}_{Ignore}^{\GlnuDSM}  
\end{eqnarray}
with the  all-loop-orders renormalized Goldstone-mode SSB $\GlnuDSM$ effective potential 
\footnote{
	It is not lost on the authors that, 
	since we derived it from {\bf connected} amputated Greens functions 
	(where all vacuum energy and disconnected vacuum bubbles 
	are absorbed into an overall phase, 
	which cancels exactly in the S-matrix \cite{Bjorken1965,ItzyksonZuber}), 
	the vacuum energy in  $L^{Eff;\GlnuDSM}_{\phi}$ in
	(\ref{LEffectiveGoldstoneTheorem}) is exactly zero.
}
\begin{eqnarray}
\label{GoldstoneWavefunction}
V^{Eff;Goldstone}_{\GlnuDSM} &=&  \lambda_{\phi}^2 \Big[ \phi^{\dagger}\phi-\half\HVEV^2 \Big]^2 \,.
\end{eqnarray} 
Eqns. (\ref{LEffectiveGoldstoneTheorem},\ref{GoldstoneWavefunction}) are the $\GlnuDSM$ effective SSB Lagrangian, 
derived from the global $SU(3)_{color}\times SU(2)_L\times U(1)_Y$ Lagrangian 
$L_{\GlnuDSM}$ in
(\ref{SMFieldsLagrangian}).
It obeys the Goldstone theorem; is minimized at $(H=\HVEV, {\vec \pi}=0)$;
obeys stationarity of that true minimum \cite{ItzyksonZuber} at $\HVEV$; 
and preserves the theory's renormalizability and unitarity, 
which require \cite{ItzyksonZuber,Lee1970,Symanzik1970a,Symanzik1970b,Vassiliev1970} that dimensionless wavefunction renormalization 
$\HVEV_{Bare}=Z_{\phi}^{1/2}\HVEV$ not attract any relevant operators.

With Goldstone mode  wavefunction renormalization
\begin{eqnarray}
\Gamma_{0,2}(;q,-q)-\Gamma_{0,2}(;00) = q^2 +{\cal O}_{Ignore}^{\nu_D SM^G}\,,
\end{eqnarray}
the {\bf coordinate-space} effective Lagrangian reads
\begin{eqnarray}
\label{GoldstoneLagrangian}
L^{Eff}_{\GlnuDSM} &=& 
	\vert \partial_{\mu}\phi \vert ^2 
	-\lambda_{\phi}^2  \Big[ {\phi}^{\dagger}\phi -\half\HVEV ^2 \Big]^2 \nonumber  \\
&+&{\cal O}_{Ignore}^{\GlnuDSM}  \\
&=&\vert \partial_{\mu}\phi \vert ^2 
	-\lambda_{\phi}^2  \Big[ \frac {(h^2+{\vec \pi}^2)}{2}+\HVEV h \Big]^2 \nonumber \\
&+&{\cal O}_{Ignore}^{\GlnuDSM} \,.\nonumber
\end{eqnarray}

We conclude section \ref{RenormalizationSM} with a few observations about equation  (\ref{GoldstoneLagrangian}):
\begin{itemize}
\item It includes all ${\cal O}(\Lambda^2),{\cal O}(\ln \Lambda^2)$ 
and finite terms that arise, 
to all perturbative loop-orders, 
in the full $SU(3)_{Color}\times SU(2)_L \times U(1)_Y$ theory,
i.e.  due to virtual  fermions and scalars. 
\item $SU(2)_L\times U(1)_Y$ is spontaneously broken.

\item The ultra-violet properties of the  Goldstone-mode $\GlnuDSM$ effective potential
	are analogous with those of the Goldstone mode of the global Schwinger \LSM \cite{Schwinger1957} 
	corresponding to the y-axis 
	of the quarter-plane characterizing the \GMLfull \LSM~ with PCAC,
	as in Figure 1 in \cite{Lynnetal2012} 
	and Figure 12-12 in the textbook by C.Itzykson \& J-C. Zuber \cite{ItzyksonZuber}. 
\item All relevant operators (e.g. UVQD$\sim {\cal O}(\Lambda^2)$ 
	in the $\GlnuDSM$ have
	vanished identically due to the Goldstone theorem.
\item The $N=M=1$ WTI
\bea
\label{BEHMassSquared}
 \Gamma_{2,0}(00;)&=&\HVEV\Gamma_{1,2}(0;00)+\Gamma_{0,2}(;00) \nonumber \\
&=&\HVEV\Gamma_{1,2}(0;00)
\eea
	relates the BEH mass-squared  from (\ref{GoldstoneLagrangian}) 
	to the coefficient of the $h{\vec \pi}^2$ vertex, so that
	\be
	\label{BEHMass}
	m_{h}^2\equiv m_{BEH}^2 = 2\lambda_{\phi}^2\HVEV^2
	\ee 
	arises entirely from SSB.
	
\item The observable BEH resonance pole-mass-squared:
\begin{eqnarray}
\label{GaugeIndependentBEHPoleMass}
m^2_{h;Pole} \nonumber &=& 2\lambda_\phi^2 \HVEV^2\Big[ 1- 2\lambda_\phi^2 \HVEV^2 
\int dm^2 \frac{\rho_{h}(m^2)}{m^2 - i\epsilon} \Big]^{-1} \nonumber \\
&+&{\cal O}_{Ignore}^{\nu_D SM^G} 
\end{eqnarray}
\item $\HVEV=Z_{\phi}^{-\half}\HVEV_{Bare}$ absorbs no relevant operators 
	(i.e. at worst $\sim \ln \Lambda^2$).  
\item As promised, $\vec{\tilde\pi}$ are true NGB. 
	In the unitary Kibble representation 
	\cite{Ramond2004,Georgi2009},
	\begin{eqnarray}
	\label{GoldstoneKibbleLagrangian}
	L^{Eff}_{\GlnuDSM} 
	&=& \frac{1}{2} \left( \partial_{\mu} {\tilde H} \right)^2 
	+ \frac{1}{4} {\tilde H}^2 Tr \left[ \partial_{\mu} U^{\dagger} \partial_{\mu} U \right] \nonumber \\
	&-&\frac{\lambda_{\phi}^2}{4}  \Big[ {\tilde H}^2 -\HVEV ^2 \Big]^2
	+{\cal O}_{Ignore}^{\nu_D SM^G} \nonumber \\
U &=& e^{   i  {\vec \sigma}  \cdot  \vec{\tilde\pi} /\HVEV   }
	\end{eqnarray}
\item	The vanishing of $\mpisq$ 
	allows ${\vec{\tilde\pi}}$ to have only derivative couplings
	and therefore possess the required shift symmetry for constant $\vec \theta$
\begin{eqnarray}
\label{ShiftSymmetryAgain} 
	{\vec {\tilde\pi}}\to {\vec {\tilde\pi}}+{\vec {\tilde\pi}}\times {\vec \theta} + \HVEV {\vec \theta}+{\cal O}(\theta^2)
\end{eqnarray}
\end{itemize}

\section{The Goldstone theorem and axial-vector WTI cause certain heavy BSM particles to de-couple from the low-energy $\GlnuDSM$ scalar-sector effective Lagrangian}
\label{RenormalizationBSM}

If the Euclidean cutoff $\Lambda^2$ were a true proxy for very heavy BSM particles, 
we'd already be in a position to comment on their de-coupling.
Unfortunately, although the literature often cites such proxy, it is simply not true. 
To quote Ergin Sezgin 
``In order to prove theorems that reveal symmetry-driven results in field theories,
one must keep {\it all} of the terms arising from {\it all} Feynman graphs,  
not just a selection of interesting terms from a representative subset of Feynman graphs."

\subsection{Criteria for the extension of axial-vector WTIs to include the $\GlnuDSM$, extended with certain heavy BSM particles}
\label{LowEnergyBSM}

In Appendix \ref{WTIproof}, 
we derive the T-Matrix and Green's function WTI for the case of the $\GlnuDSM$ Lagrangian, 
extended to include neutrino Dirac masses.

We here derive criteria that anomaly-free Beyond the Standard Model (BSM) spin $S=0$ scalars $\Phi$, and $S=\half$ fermions $\psi$, must obey 
in order that the axial-vector WTI remain true. 

{\bf 1) Begin by focussing on the global $SU(2)_L$ $\GlnuDSM$ isospin current}
and dividing it into vector and axial-vector parts:
\begin{eqnarray}
\label{TotalCurrent}
2 {\vec J}^{\mu}_{L+R;\GlnuDSM}&=&{\vec \pi}
\times \partial^\mu {\vec \pi} \nonumber \\
 &+&  \sum_{c} {\bar q}^c \gamma^\mu {\vec t} q^c+  {\bar l} \gamma^\mu {\vec t} l \nonumber \\
2 {\vec J}^{\mu}_{L-R;\GlnuDSM}&=& {\vec \pi} \partial^\mu H-H\partial^\mu {\vec \pi} \nonumber \\
&+&  \sum_{c} {\bar q}^c \gamma^\mu\gamma^5 {\vec t} q^c+  {\bar l} \gamma^\mu \gamma^5{\vec t} l \nonumber \\
 {\vec J}^{\mu}_{L;\GlnuDSM}&=& {\vec J}^{\mu}_{L+R;\GlnuDSM}+ {\vec J}^{\mu}_{L-R;\GlnuDSM}\quad
\end{eqnarray}
with colors $c=r,w,b$, isospin ${\vec t}=\half {\vec \sigma}$, Pauli matrices $\vec \sigma$.

The classical equations of motion show only that the $SU(2)_L$ isospin current is conserved  
\bea
\label{DivergenceSMCurrent}
\partial_{\mu} {\vec J}^{\mu}_{L;\GlnuDSM}&=&0 
\eea
But in the $\GlnuSM$ studied here, $CP$ is conserved, so that on-shell and off-shell connected amputated 
T-matrix elements and Green's functions of an odd number of $\vec \pi$s 
and their derivatives, are zero. 
They also vanish for an odd number of $\vec \pi$s 
and fermion bi-linears with the isospin quantum numbers of $\vec \pi$.

{ 
$SU(2)_{L-R}$ is not a sub-group of the $SU(2)_L$ symmetry group, 
but CP conservation ensures that the global vector current 
transforms as an even number of $\vec \pi$s, 
while the global axial-vector current transforms as an odd number of $\vec \pi$s. 
Thus, for M even,
}
\begin{eqnarray}
\label{AxialCurrentMEven}
&&\Big< 0\vert T\Big[ \Big(  {\vec J}^{\mu}_{L-R;\GlnuDSM} (z)\Big)    \\
&&\quad \quad \times h(x_1)...h(x_N) \pi^{t_1}(y_1)...\pi^{t_M}(y_M)\Big]\vert 0\Big>_{Connected}^{M~even} =0 
\nonumber \\
&& \Big< 0\vert T\Big[ \Big( \partial_{\mu} {\vec J}^{\mu}_{L-R;\GlnuDSM}(z)\Big)  \nonumber \\
&&\quad \quad \times h(x_1)...h(x_N)  \pi^{t_1}(y_1)...\pi^{t_M}(y_M)\Big]\vert 0\Big>_{Connected}^{M~even}=0\,.
\nonumber
\end{eqnarray}
Meanwhile, (\ref{TotalCurrent},\ref{DivergenceSMCurrent}) show that, for M odd,
\begin{eqnarray}
\label{AxialCurrentMOdd}
&& \Big< 0\vert T\Big[ \partial_{\mu} \Big(  {\vec J}^{\mu}_{L-R;\GlnuDSM} (z)\Big)   \nonumber \\
&&\quad \quad \times h(x_1)...h(x_N) \pi^{t_1}(y_1)...\pi^{t_M}(y_M)\Big]\vert 0\Big>_{Connected}^{M~odd}  \nonumber \\
&& =\Big< 0\vert T\Big[ \partial_{\mu} \Big({\vec J}^{\mu}_{L;\GlnuDSM}-  {\vec J}^{\mu}_{L+R;\GlnuDSM}\Big) (z)  \nonumber \\
&&\quad \quad \times h(x_1)...h(x_N) \pi^{t_1}(y_1)...\pi^{t_M}(y_M)\Big]\vert 0\Big>_{Connected}^{M~odd}  \nonumber \\
&&=\Big< 0\vert T\Big[ \Big( \partial_{\mu} {\vec J}^{\mu}_{L;\GlnuDSM}(z)\Big)  \nonumber \\
&&\quad \quad \times h(x_1)...h(x_N)\pi^{t_1}(y_1)...\pi^{t_M}(y_M)\Big]\vert 0\Big>_{Connected}^{M~odd} \nonumber \\
&&=0 \,.
\end{eqnarray}
{ Thus the $SU(2)_{L-R}$ current is ``effectively conserved.'' } 

Similarly, although $SU(2)_{L+R}$ is not a sub-group of $SU(2)_L$, 
its current is also effectively conserved 
for Green's functions and T-Matrix elements for all M:
\begin{eqnarray}
\label{VectorCurrentConserved}
&& \Big< 0\vert T\Big[ \Big( \partial_{\mu} {\vec J}^{\mu}_{L+R;\GlnuDSM}(z)\Big)   \\
&&\quad \times h(x_1)...h(x_N)  \pi^{t_1}(y_1)...\pi^{t_M}(y_M)\Big]\vert 0\Big>_{Connected}=0
\nonumber
\end{eqnarray}

This paper is based on the effective conservation 
of $ {\vec J}^{\mu}_{L-R;\GlnuDSM}$ 
for on-shell and off-shell connected amputated  Green's functions and T-matrix elements, 
in (\ref{AxialCurrentMEven},\ref{AxialCurrentMOdd})
\footnote{
{ Had CP not been conserved, the vector and axial-vector currents would not have
separately been conserved. We look forward to future works to exploring the consequences of 
soft CP violation on WTIs.}
}.

{\bf 2) We extend the $\GlnuDSM$ with certain BSM matter particles.} These must carry zero anomaly.

In order to force renormalized connected amplitudes with an odd number of $\pi$s to vanish, the new particles $\Phi,\psi$ are taken in this paper to conserve $CP$.
Divide the conserved isospin current into axial-vector (i.e. transforming as an odd number of $\vec \pi$s under isospin) and vector (i.e. transforming as an even number of $\vec \pi$s under isospin) parts. 
\begin{eqnarray}
\label{DivergenceTotalCurrent}
\partial_{\mu} {\vec J}^{\mu}_{L;Total} \equiv\partial_{\mu} \Big( {\vec J}^{\mu}_{L;\GlnuDSM} + {\vec J}^{\mu}_{L;BSM} \Big)&=&0 \nonumber \\
 {\vec J}^{\mu}_{L+R;BSM} + {\vec J}^{\mu}_{L-R;BSM} &\equiv& {\vec J}^{\mu}_{L;BSM}  \nonumber \\
 {\vec J}^{\mu}_{L-R;\nu_D SM^G} + {\vec J}^{\mu}_{L-R;BSM} &\equiv& {\vec J}^{\mu}_{L-R;Total} \nonumber \\
\Big< 0\vert T\Big[ \Big( \partial_{\mu} {\vec J}^{\mu}_{L-R;Total}(z)\Big) h(x_1)...h(x_N)    \nonumber \\
 \quad \times \pi^{t_1}(y_1)...\pi^{t_M}(y_M)\Big]\vert 0\Big>_{Connected}&=&0
\end{eqnarray}

{\bf 3)  Canonical quantization is imposed on the exact renormalized fields,} 
yielding equal-time quantum commutators at space-time points $y, z$. 
The BSM axial-vector currents 
must commute with $H$ and ${\vec \pi}$:
\begin{eqnarray}
\label{eqn:PCACEqTimeCommBSM}
 \delta(z_0-y_0)\left[ {\vec J}^0_{L-R;BSM}(z),H(y)\right] &=& 0 \\
 \delta(z_0-y_0)\left[ J_{L-R;BSM}^{0;i}(z),\pi^j (y)\right] &=& 0\,. \nonumber 
\end{eqnarray} 
Only certain BSM matter will obey this condition.

{\bf 4) BSM scalars must have zero VEV.}
Only certain BSM matter will obey this condition.
Note that Green's functions are then usually 1-BSM Scalar-Reducible, by cutting a BSM-Scalar line.

{\bf 5) Certain surface integrals must vanish:} 
Appendix A used pion-pole dominance to derive 1-soft-pion theorems, 
which require that the connected surface integral (\ref{SurfacePionPoleDominance}) vanish.
In (\ref{SurfacePionPoleDominance})  
we have N external renormalized $h=H-\HVEV$ (coordinates x, momenta p), 
M external ($CP=-1$) renormalized ${\vec \pi}$ (coordinates y, momenta q, isospin t).
Because $CP$ is conserved, only axial-vector WTI are needed 
to put the effective Lagrangian into the desired form.
We form the surface integral
\begin{eqnarray}
\label{SurfacePionPoleDominance}
&&\lim_{k_\lambda \to 0} \int d^4z e^{ikz} \partial_{\mu} \Big< 0\vert T\Big[ \Big( 2 {\vec J}^{\mu}_{L-R;Total} +\HVEV \partial^\mu {\vec \pi}\Big)(z)  \nonumber \\
&&\quad \quad \times h(x_1)...h(x_N) \pi^{t_1}(y_1)...\pi^{t_M}(y_M)\Big]\vert 0\Big>_{Connected} \nonumber \\
&& =\int d^4z \partial_{\mu} \Big< 0\vert T\Big[ \Big( 2 {\vec J}^{\mu}_{L-R;Total}+\HVEV \partial^\mu {\vec \pi}  \Big)(z)  \nonumber \\
&&\quad \quad \times h(x_1)...h(x_N) \pi^{t_1}(y_1)...\pi^{t_M}(y_M)\Big]\vert 0\Big>_{Connected} \nonumber \\
&& =\int_{3-surface}  d^3z \quad {\widehat {z}_\mu}^{3-surface} \nonumber \\
&&\quad \quad \times \Big< 0\vert T\Big[ \Big( 2 {\vec J}^{\mu}_{L-R;Total}+\HVEV \partial^\mu {\vec \pi} \Big)(z^{3-surface}\to \infty)  \nonumber \\
&&\quad \quad \times h(x_1)...h(x_N) \pi^{t_1}(y_1)...\pi^{t_M}(y_M)\Big]\vert 0\Big>_{Connected} \nonumber \\
&& \quad \quad =0 \,,
\end{eqnarray}
where we have used Stokes theorem, 
and $ {\widehat {z}_\mu}^{3-surface}$ is a unit vector normal to the $3-surface$. 
The time-ordered-product constrains the $3-surface$ to lie on, or inside, the light-cone. 

At a given point on the surface of a large enough 4-volume $\int d^4z$ (i.e. the volume of all space-time): all fields are asymptotic in-states and out-states, properly quantized as free fields, with each field species orthogonal to the others,
and  they are evaluated at equal times, making time-ordering un-necessary at $(z^{3-surface}\to \infty)$. 
Input the global axial-vector current (\ref{TotalCurrent}) to  (\ref{SurfacePionPoleDominance}), using $\partial_\mu \HVEV =0$. 
The contribution to (\ref{SurfacePionPoleDominance}) from $\GlnuDSM$ vanishes
\begin{eqnarray}
\label{SurfaceIntegralA}
&& \int_{3-surface}  d^3z \quad {\widehat {z}_\mu}^{3-surface} \Big< 0\vert T\Big[  \nonumber \\
&&\quad  \times \Big(  {\vec \pi} \partial^\mu h-h\partial^\mu {\vec \pi} + \sum_{c} {\bar q}^c\gamma^\mu \gamma^5  {\vec t} q^c\nonumber \\
&&\quad  \quad \quad \quad  + {\bar l}\gamma^\mu \gamma^5 {\vec t} l \Big)(z^{3-surface}\to \infty)  \nonumber \\
&&\quad  \times h(x_1)...h(x_N) \pi^{t_1}(y_1)...\pi^{t_M}(y_M)\Big]\vert 0\Big>_{Connected} \nonumber \\
&&\quad  =0
\end{eqnarray}
The 1st and 2nd terms vanish because the BEH $h$ is massive. 
The 3rd term vanishes because all quarks have non-zero Dirac masses.
The 4th term vanishes because all leptons in the $\GlnuDSM$, {\bf including neutrinos}, have non-zero Dirac masses.
Propagators connecting massive
$h, q_{L}^c, l_{L} $ 
from points on $z^{3-surface}\to \infty$ to the localized interaction points $(x_1...x_N;y_1...y_M)$, must stay inside the light-cone, die off exponentially with mass,
and are incapable of carrying information that far.

It is the central observation for ``pion-pole-dominance"  and this paper, that {\em this argument fails} for the remaining term 
in the axial-vector current $2{\vec J}^{\mu}_{L-R;\GlnuDSM}$ in (\ref{TotalCurrent}).
\begin{eqnarray}
\label{NGBSurfaceIntegral}
&& \int_{2-surface}  d^2z \quad {\widehat {z}_\mu}^{2-surface}  \nonumber \\
&&\quad \quad \times \Big< 0\vert T\Big[  \Big(-\HVEV \partial^\mu {\vec \pi} \Big)(z^{2-surface}\to \infty)  \nonumber \\
&&\quad \quad \times h(x_1)...h(x_N) \pi^{t_1}(y_1)...\pi^{t_M}(y_M)\Big]\vert 0\Big>_{Connected} \nonumber \\
&&\quad \quad \neq 0
\end{eqnarray}
$\vec \pi$ is massless, capable of carrying (along the light-cone) long-ranged pseudo-scalar forces out to the $2-surface$  $(z^{2-surface}\to \infty)$: i.e. the very ends of the light-cone (but not inside it).
That massless-ness is the basis of our pion-pole-dominance-based axial-vector WTIs which, as derived in Appendix A, give 1-soft-pion theorems, 
 infra-red finiteness for $\mpisq =0$,  and a ``Goldstone theorem.''

{\bf 6) In order to include spin $S=0$ scalar, and $S=\half$ fermionic, BSM matter representations} in our axial-vector Ward-Takahashi identities, a certain surface integral  must vanish.
\begin{eqnarray}
\label{SurfaceIntegralB}
&&\int d^4z \partial_{\mu} \big< 0\vert T \Big[ \Big( {\vec J}^{\mu}_{L-R;BSM}(z) 
\Big) \nonumber \\
&& \quad \quad \quad \quad \times h(x_1)...h(x_N)\pi_{t_1}(y_1)...\pi_{t_M}(y_M) \Big]\vert 0 \big>  \nonumber \\
&& = 0 
\end{eqnarray}
Additional BSM particles must generically be  massive, 
and thus incapable of carrying information to the surface at infinity. They must also have zero vacuum expectation values.
Only certain BSM matter will obey this condition.

We examine below the consequences of extending the $\GlnuDSM$ to include certain 
high-mass-scale $M_{Heavy}^2\gg m_{Weak}^2$ BSM matter, 
especially the relevant operator contributions  ${\cal O}(\Lambda^2)$, 
${\cal O}(M_{Heavy}^2 \ln \Lambda^2)$, ${\cal O}(M_{Heavy}^2 \ln M_{Heavy}^2)$,  ${\cal O}(M^2_{Heavy})$,
 ${\cal O}(M_{Heavy}^2 \ln m_{Weak}^2)$ 
and ${\cal O}(m_{Weak}^2 \ln M_{Heavy}^2)$,  
to the effective Lagrangian of weak=scale $\GlnuDSM$ scalars $\phi$. 
We show that, for low-energy $\left|q^2\right|,m_{Weak}^2  \ll M_{Heavy}^2$ physics, 
the heavy degrees of freedom decouple completely
\footnote{
Except  for high-precision electro-weak T and U \cite{Kennedy1988, 
PeskinTakeuchi, Ramond2004}.
}, 
including marginal operators $\sim {\cal O}( \ln M_{Heavy}^{2})$, 
leaving only irrelevant operators, at worst $\sim {\cal O}( M_{Heavy}^{-2})$.
We demonstrate this below for two heavy BSM particle examples, a heavy fermion and a heavy scalar.

\subsection{$\nu {\mathrm {MSM}}^{\rm G}_{tb\tau\nu_\tau}$: Singlet right-handed Type I See-saw Majorana neutrino $\nu_R$ 
with $M_{\nu_R}^2\gg m_{BEH}^2$}
\label{HeavyNeutrino}

For the heavy fermion we consider a global $SU(3)_C\times SU(2)_L\times U(1)_Y$ singlet 
right-handed Majorana neutrino $\nu_R$, with $M_{\nu_R}^2\gg m_{Weak}^2$, 
such as might be involved in a Type 1 See-Saw with a left-handed neutrino $\nu_L$, 
with Yukawa coupling $y_{\nu}$ and resulting Dirac mass $m_D=y_{\nu}\HVEV /\sqrt{2}$. 
We add to the renormalized theory
\begin{eqnarray}
\label{MajoranaNeutrino}
L^{Majorana}_{\nu_R}&=& -M_{\nu_R} ({\nu_R}{\nu_R}+{\bar \nu}_R {\bar \nu}_R)/2
\end{eqnarray}

Since $\nu_R$ is a $SU(2)_L$  singlet, its currents
\begin{eqnarray}
\label{MajoranaCurrent}
{\vec J}^{\mu;Majorana}_{L;{\nu_R}}={\vec J}^{\mu;Majorana}_{L+R;{\nu_R}}={\vec J}^{\mu;Majorana}_{L-R;{\nu_R}}=0
\end{eqnarray}
satisfy all of the criteria, in Sub-section \ref{LowEnergyBSM},
for the extension of our axial-vector Ward-Takahashi IDs:
\begin{eqnarray}
\label{CriteriaMajoranaCurrent}
&&\Big< 0\vert T\Big[  \partial_{\mu}\Big( {\vec J}^{\mu}_{L-R;\GlnuDSM}+{\vec J}^{\mu;Majorana}_{L-R;\nu_R}\Big)(z)  \nonumber \\
&&\quad \times h(x_1)...h(x_N)  \pi^{t_1}(y_1)...\pi^{t_M}(y_M)\Big]\vert 0\Big>_{Connected}=0
\nonumber \\
&& \delta(z_0-y_0)\left[ {\vec J}^{0;Majorana}_{L-R;\nu_R}(z), H(y)\right] = 0 \nonumber \\
&& \delta(z_0-y_0)\left[ {\vec J}^{0;Majorana}_{L-R;\nu_R}(z), {\vec \pi}(y)\right] = 0 \nonumber \\
&&\int d^4z \partial_{\mu} \big< 0\vert T \Big[ \Big( {\vec J}^{\mu;Majorana}_{L-R;\nu_R}(z) 
\Big) \nonumber \\
&& \quad \quad \cdot h(x_1)...h(x_N)\pi_{t_1}(y_1)...\pi_{t_M}(y_M) \Big]\vert 0 \big> =0 
\end{eqnarray}

Since it is massive $\nu_R$ cannot carry information to the surface of the 4-volume $\int d^4 z$, nor can it induce any ``neutrino-pole-dominance" terms.
It follows that the WTI for T-Matrix elements (\ref{TMatrixID}), 
Greens functions (\ref{eqn:BWLeeWTI}),
Adler's self-consistency and IR finiteness (\ref{Adler}),
and Goldstone theorem (\ref{GFGoldstoneTheorem}),
are still true for the $\GlnuDSM$ with a non-zero Majorana neutrino mass.

In order that total neutrino masses $\sim m^{2;Dirac}_\nu /M_{\nu_R}$ remain non-zero in this Type I See-saw, we can't take the strict $M_{\nu_R} \to \infty$ limit: that would destroy our axial-vector WTIs.  Instead we take $1\gg m_{Weak}^2/M_{\nu_R}^2 > 0$, so that $\nu_R$ decouples {\bf in practice}.

The weak-scale effective Lagrangian therefore remains 
(\ref{LEffectiveGoldstoneTheorem},\ref{GoldstoneWavefunction}).

\subsection{Singlet $M^2_S \gg m_{BEH}^2$ real scalar field $S$ with discrete $Z_2$ symmetry and $\SVEV=0$}
\label{HeavyScalar}

Consider  an $SU(3)_C\times SU(2)_L\times U(1)_Y$ singlet  real scalar $S$, 
with  ($S\to-S$) $Z_2$ symmetry,
$M_S^2\gg m_h^2$, and $\SVEV =0$. 
We add to the renormalized theory
\begin{eqnarray}
\label{SingletScalarLagrangian}
L_S&=&\half(\partial_{\mu}S)^2 -V_{\phi S} \\
V_{\phi S} &=& \half M_S^2 S^2 + \frac {\lambda_S^2}{4} S^4 + \half \lambda_{\phi S}^2 S^2 \left[ \phi^\dagger\phi -\half \HVEV^2 \right] \nonumber
\end{eqnarray}
with $M_S^2>0$. Again, 
\begin{eqnarray}
\label{SingletCurrent}
{\vec J}^{\mu}_{L;S}={\vec J}^{\mu}_{L+R;S}={\vec J}^{\mu}_{L-R;S}=0
\end{eqnarray}
and all the analogues of equation (\ref{CriteriaMajoranaCurrent}) follow.

Since it is massive,
$S$ cannot carry information to the surface of the 4-volume $\int d^4 z$. 
$S-h$ mixing, 
which might well have spoiled the protection that the WTI provide to $m_h^2$, 
is forbidden by the  $Z_2$ symmetry $S\to-S$.
It follows that the axial-vector WTI for T-Matrix elements (\ref{TMatrixID}), 
Greens functions (\ref{eqn:BWLeeWTI}),
Adler's self-consistency and IR finiteness (\ref{Adler}),
including the Goldstone theorem, 
are still true for the $\GlnuDSM$ extended to include this scalar singlet.
Note that Green's functions are usually 1-$S$-Reducible, by cutting an $S$ line.

In the $m_{Weak}^2/M_S^2 \to 0$ limit, the weak-scale effective Lagrangian therefore  
again remains (\ref{LEffectiveGoldstoneTheorem},\ref{GoldstoneWavefunction}).  

\section{Conclusion: Historically, complete decoupling of heavy invisible particles is the usual physics experience}
\label{Conclusions}

We defined the $\GlnuDSM$ 
as the global $SU(3)_C\times SU(2)_L\times U(1)_Y$ model 
of a complex Higgs doublet and 3rd generation SM quarks and leptons, 
augmented by a right-handed neutrino with Dirac mass.
With SM isospin and hypercharge assignments for fermions, 
$\GlnuDSM$ has zero axial anomaly.
We showed that, { in the presence of CP conservation,} 
the weak-scale low-energy effective Lagrangian 
of the spontaneously broken  $\GlnuDSM$ 
is severely constrained by, and  protected by, 
new  rigid/global SSB axial-vector Ward-Takahashi identities (WTI)
including an equivalent of the  Goldstone theorem.
In particular, 
the weak-scale SSB $\GlnuDSM$ has an $SU(2)_L$  shift symmetry
for constant $\vec \theta$
\begin{eqnarray}
\label{HiddenshiftSymmetryPrime}
{\vec {\tilde \pi}} \to {\vec {\tilde \pi}} + {\vec {\tilde \pi}}\times {\vec \theta} +\HVEV {\vec \theta} + {\cal O}(\theta^2)\,.
\end{eqnarray}
This protects it,
and  causes the  complete  decoupling 
of certain heavy $M^2_{Heavy} \gg m_{Weak}^2$ BSM matter-particles.
(Note that such decoupling is 
modulo special cases: e.g. heavy Majorana $\nu_R$, and possibly 
${\cal O}_{\nu_D SM^G}^{Dim \leq 4;NonAnalytic;Heavy}$, 
which are dimension$\leq4$ operators, 
non-analytic in momenta or a renormalization scale $\mu^2$, 
involve heavy particles, and are beyond the scope of this paper.)

Such heavy-particle decoupling is historically 
the usual physics experience at each energy scale 
as experiments probed smaller and smaller distances. 
After all, Willis Lamb did not need to know 
the top quark or BEH mass \cite{AppelquistCarazzone} 
in order to interpret theoretically 
the experimentally observed ${\cal O}(m_e \alpha^5 \ln \alpha)$ splitting 
in the spectrum of hydrogen. 

Such heavy-particle decoupling may be the reason why the Standard Model, viewed as an effective low-energy weak-scale theory, is the most experimentally and observationally successfull 
and accurate theory of Nature known to humans, i.e. when augmented by classical General Relativity and neutrino mixing: that ``Core  Theory"
\cite{WilcekCoreTheory} has no known experimental or observational counter-examples.

{\bf Acknowledgements}
The importance and influence of Raymond Stora's contribution to this paper
(revelatory conversations, 
correction of the authors' errors and wrong-headedness, 
illumination of the history of renormalization, 
attention/obedience to the detailed technology of renormalization, etc.) 
cannot be over-estimated. 

BWL thanks Jon Butterworth and University College London for support as an Honorary Senior Research Associate; 
Albrecht Karle and U Wisconsin at Madison for hospitality during the academic year 2014-2015; and Chris Pope, the George and Cynthia Woods Mitchell Center for Fundamental Physics and Astronomy,  and Texas A\&M University for support/hospitality, during the academic year 2010-2011, where this work began.
GDS is partially supported by CWRU grant DOE-SC0009946  
and thanks the CERN theory group for hospitality during 2012-13 when the 
groundwork for this paper was laid.

\appendix
\section{Proof of the Ward-Takahashi Identities 
\label{WTIproof}
for $SU(2)_L\times SU(2)_R$ and $SU(2)_L\times U(1)_Y$}
In 1970, B. Lee presented a series of lectures at the Cargese  Summer School
on Chiral Dynamics, with detailed results on the
renormalization of the Gell Mann-Levy Model -- the $SU(2)_L\times SU(2)_R$ Linear Sigma Model
(\LSM) with an approximate $SU(2)_{L-R}$ chiral symmetry, but with an explicit breaking term 
(known as the Partially Conserved Axial Current (PCAC), or Polkinghorne term).
Of specific interest to us, in section V of those lectures, he proved a 
tower of $SU(2)_{L-R}$ Ward-Takahashi identities (WTI) among $(h,{\vec \pi})$Scalar-sector ($\Phi$-sector)
connected amputated Green's functions, Adler's self-consistency conditions, and the Goldstone theorem..

Those WTI, and the proof thereof, 
are of immense value to us in this and companion papers.  Unfortunately,
the volume in which the lecture appears \cite{Lee1970} is difficult to obtain. We therefore
present in this Appendix, for the benefit of the reader,
those WTI and Lee's proof of them for the case of conserved axial-vector currents.  We hew closely to Lee's presentation, language, notation and pedagogy.
Although we sometimes comment/elaborate on specific details, we mostly just let Lee explain.
Because we are interested in weak interactions rather than strong interactions,
we set the explicit PCAC $SU(2)_{L-R}$ breaking term (parametrized by $\gamma$ in Lee's notation) to zero:
the result is the $SU(2)_L\times SU(2)_R$ Schwinger model \cite{Schwinger1957}.
Because we are interested in including SM fermions, whose Yukawa couplings  
break global $SU(2)_L\times SU(2)_R$
explicitly to global $SU(2)_L\times U(1)_Y$, 
we derive $SU(3)_C\times SU(2)_L\times U(1)_Y$ WTIs, analogous with those of Lee.

The  conserved vector and axial-vector currents of the $\Phi$-sector
$SU(2)_L\times SU(2)_R$ \LSM~
with the Lagrangian (\ref{eqn:VGML_Ren}) are
\begin{eqnarray}
		 \label{eqn:SchwingerCurrents}
		{\vec V}_{\mu}&=&
			{\vec \pi}\times \partial_{\mu}{\vec \pi} ;\qquad \partial^{\mu}{\vec V}_{\mu}(x) = 0 \\
		{\vec A}_{\mu}&=&
			{\vec \pi}\partial_{\mu}H - H\partial_{\mu}{\vec \pi}; \qquad \partial^{\mu}{\vec A}_{\mu}(x) = 0 \nonumber
\end{eqnarray}
In Lee's lectures, 
there is an explicit PCAC breaking of the chiral symmetry,
$\partial_{\mu}{\vec A}^{\mu}(x) = \gamma {\vec \pi}(x)$. 
In this paper we take $\gamma\equiv 0$.

In $SU(2)_L\times U(1)_Y$, we are interested in the left-handed 
combination of ${\vec V}_{\mu}$ and ${\vec A}_{\mu}$: 
$2 {\vec J}_L^\mu \equiv {\vec V}^\mu+{\vec A}^\mu$.
The addition of fermions in $SU(3)_C\times SU(2)_L\times U(1)_Y$ representations adds contributions to both ${\vec V}_{\mu}$ and ${\vec A}_{\mu}$:
The various $SU(2)_L,SU(2)_{L+R},SU(2)_{L-R}$ currents in the $\GlnuDSM$, together with their conservation laws, are given in Eqns. (\ref{TotalCurrent}) through (\ref{VectorCurrentConserved}).
We focus the remainder of our attention
on the $SU(3)_C\times SU(2)_L\times U(1)_Y$  case, which has some additional subtleties compared to Lee's $SU(2)_L\times SU(2)_R$.

We examine time-ordered amplitudes of
products of the {\bf axial-vector current} ${\vec  J}^{L-R}_{\mu}$, 
with N scalars (coordinates x, momenta p), 
and M pseudo-scalars (coordinated y, momenta q, isospin t):
$\big< 0 \vert T\Big[ 
{\vec  J}^{L-R}_\mu(z)h(x_1)\cdots h(x_N) 
						\pi^{t_1}(y_1)\cdots \pi^{t_M}(y_M)
\Big]\vert 0\big>$
Here $h=H-\HVEV$ and ${\vec \pi}$ are all-loop-orders renormalized fields, normalized so that
$\left\langle0\vert h(0) \vert h\right\rangle=1$ and
$\left\langle0\vert\pi^i(0)\vert\pi^j\right\rangle=\delta^{ij}$
We want the divergence of such amplitudes.\tabularnewline

{ 
We are reminded that, as discussed in section \ref{LowEnergyBSM}, 
such axial-vector-current amplitudes (as distinct from left-handed current amplitudes)
are of interest, and can be considered separately from vector-current amplitudes,
solely because of  CP conservation.}
Making use of axial-vector-current conservation 
(equations (\ref{TotalCurrent}) through (\ref{VectorCurrentConserved})), 
and the equal-time commutation relations { (again assuming CP conservation)} 
\bea
\delta(z_0-x_0) \left[ 2{ J}_0^{L-R;i}(z)  ,h(x)\right] &=& -i \pi^i(x)\delta^{(4)}(x-z)\nonumber\\
\delta(z_0-y_0) \left[  2{ J}_0^{L-R;i} (z),\pi^j(y)\right] &=& +i\delta^{ij} h(y)\delta^{(4)}(y-z) \quad\quad 
\eea
a short calculation reveals
\begin{widetext}
\bea
\label{eq:divAGFreal}
\partial^\mu&&  \big< 0\vert 
	T\left[2J^{L-R;t}_\mu(z)h(x_1)\cdots h(x_N)\pi^{t_1}(y_1)\cdots \pi^{t_M}(y_M)\right]
		\vert 0 \big>  \nonumber\\
	&=&
	i \sum_m^M \big< 0\vert 
		T\left[ h(x_1)\cdots h(x_N)h(z)
		\pi^{t_1}(y_1)\cdots \widehat{ {\pi}^{t_m}(y_m) }\cdots\pi^{t_M}(y_M)
		\right]
		\vert 0 \big> \delta^{(4)}(y_m-z) \delta^{t,t_m} \nonumber \\
	&+&
	i\HVEV \sum_m^M \big< 0\vert 
		T\left[ h(x_1)\cdots h(x_N)
		\pi^{t_1}(y_1)\cdots \widehat{{ \pi}^{t_m}(y_m)}\cdots\pi^{t_M}(y_M)
		\right]
		\vert 0 \big> \delta^{(4)}(y_m-z) \delta^{i,i_j} \nonumber\\
	&-& 
	i \sum_n^N \big< 0\vert 
		T\left[ h(x_1)\cdots \widehat{  h(x_n)}\cdots h(x_N)
		\pi^{t}(x_n)\pi^{t_1}(y_1)\cdots\pi^{t_M}(y_M)
		\right]
		\vert 0 \big>  \delta^{(4)}(x_n-z)
\eea
\end{widetext}
where $\widehat {\pi^{t_m}(y_m)}$ or $\widehat {h(x_n)}$
indicates that that copy of $\pi$ or $h$ is to be omitted from the product of fields.
The LHS of (\ref{eq:divAGFreal}) has $M$ pion fields. 
On the RHS, the first 2 terms  have $M-1$, and  the 3rd term $M+1$, pions.

The fermion contributions to ${\vec J}^{L-R}_\mu$ 
commute with $h(x)$ and $\pi(y)$ and so do not contribute to the RHS of 
(\ref{eq:divAGFreal}). Fermion  contributions to the LHS remain.

We define the Fourier Transform of these amplitudes in the usual way
\begin{widetext}
\bea
\label{eq:GFFT}
	iG_\mu^{t;t_1\cdots t_M}&&\left(k;p_1\cdots p_N;q_1\cdots q_M\right) 
		\left(2\pi\right)^2 \delta^{(4)}\left(k + \sum_n^N p_n +\sum_m^M q_m\right)\\
	&&
	\equiv
	\int d^4z e^{i k \cdot z}
	\Pi_{n=1}^N\int d^4x_n e^{i p_n \cdot x_n}
	\Pi_{m=1}^M\int d^4y_m e^{i q_m \cdot y_m}
	\big< 0\vert
		T\left[2 J^{L-R;t}_\mu(z) h(x_1)\cdots h(x_N) \pi^{t_1}(y_1)\cdots\pi^{t_M}(y_M)
		\right]
	\vert 0 \big> \nonumber
\eea
\end{widetext}

To economize on notation, going forward we will omit all isospin indices, 
letting momenta stand in for the isospin indices as well.  

The reader is warned that, in this Appendix (and in Lee), it is assumed that $\vec \pi$ are the only massless fields in the theory, so that certain surface integrals, which are discussed in the body of this paper, vanish.

Taking the  Fourier Transform of the divergence of the amplitude, 
and applying Stokes theorem:
\begin{widetext}
\bea
\int d^4z e^{i k\cdot z} 
\Pi_{n=1}^N \int d^4x_i e^{i p_n\cdot x_n} 
\Pi_{m=1}^M \int d^4y_m e^{i q_j\cdot y_m}
\partial_z^\mu \big< 0\vert
		T\left[2 J^{L}_\mu(z) h(x_1)\cdots h(x_N) \pi(y_1)\cdots\pi(y_M)
		\right]
	\vert 0 \big> 
&=&\nonumber\\
k^\mu G_\mu(k;p_1\cdots p_N; q_1\cdots q_M) 
\left(2\pi\right)^4 \delta^{(4)}\left(k + \sum_n^N p_n + \sum_m^N q_m\right)
\eea
\end{widetext}
\bea
\label{eq:divergenceofG}
&& k^\mu \quad \!\!\!\!\!\!G_\mu(k;p_1\cdots p_N; q_1\cdots q_M)   \\ 
&& =\sum_n^M G(p_1\cdots {\widehat {p_m}} \cdots p_N; k+p_m, q_1\cdots q_M)  \nonumber\\
&& -\sum_m^M G(k+q_m,p_1\cdots p_N;q_1\cdots {\widehat{q_m}}\cdots q_M) \delta_{t,t_m}\nonumber \\
&& -\HVEV \sum_m^M G(p_1\cdots p_N;q_1\cdots {\widehat {{q}_m} }\cdots q_M) \nonumber\\
&&\quad\quad\quad\quad\quad\times
\delta_{t,t_m}  \left(2\pi\right)^4 \delta^{(4)}(k+q_m) \nonumber
\eea

This holds for all $N,M\geq1$. For $N=0$ and $M=1$
\be
\label{eq:divergenceofG01}
k^\mu G_\mu(k;;q) = i\HVEV
\ee

The T-Matrix restricts us to {\bf connected} graphs.  
The last term on the RHS of (\ref{eq:divergenceofG}) corresponds entirely to disconnected graphs.
Denote by $H_\mu$ and $H$ the connected parts of the amplitudes $G_\mu$ and $G$ defined above.
Then 
\bea
\label{eq:divergenceofH}
k^\mu &&\!\!\!\!\!\!H_\mu(k;p_1\cdots p_N; q_1\cdots q_M) \nonumber \\
&=& \sum_n^N H(p_1\cdots {\widehat {{p}_n}} \cdots p_N; k+p_n, q_1\cdots q_M)\\
&&- \sum_m^M H(k+q_m,p_1\cdots p_N;q_1\cdots {\widehat {{q}_m}}\cdots q_M) \delta_{t,t_m}\nonumber
\eea
for all $N,M\geq1$. For $N=0$ and $M=1$
\be
\label{eq:divergenceofH01}
k^\mu H_\mu(k;;q) = i\HVEV\,.
\ee

In order to derive ``1-soft-pion theorems", the limit $k\to0$ is taken in equations (\ref{eq:divergenceofH}) 
and (\ref{eq:divergenceofH01}).  Here we must be careful.  Because of the
pion pole in $H_\mu$ at $k^2=0$, $k^\mu H_\mu \to constant$.  We therefore
isolate the pion-pole contribution to $H_\mu$ by writing
\begin{flalign}
\quad H_\mu&(k;p_1\cdots p_N;q_1\cdots q_M)\nonumber&&\\
&\equiv i\HVEV k_\mu H(k;p_1\cdots p_N;q_1\cdots q_M)&&\\
&+ {\bar H}_\mu(k;p_1\cdots p_N;q_1\cdots q_M)\nonumber&&
\end{flalign}
The first term contains the pion-pole contribution; 
the second term is non-singular at $k^2=0$.

With this new decomposition, the LHS of equation (\ref{eq:divergenceofH}) is
\be
i\HVEV k^2 H(p_1\cdots p_N; k q_1 \cdots q_M) + k^\mu {\bar H}_\mu\,.
\ee
As $k\to 0$, the second term vanishes, and the first term goes to a limit,
since H has a pole at $k^2=0$.  Therefore
\bea
\label{eq:WTIa}
i\HVEV && \lim_{k\to0} k^2H\left(p_1\cdots  p_N; k q_1 \cdots q_M\right)\\
 &=& \sum_n^N H\left(p_1\cdots {\widehat {p_n}}\cdots p_N; p_n, q_1 \cdots q_M\right)\nonumber\\
&&- \sum_m^M H\left(p_1\cdots  p_N q_m ; q_1 \cdots {\widehat {q_m}} \cdots q_M\right) \delta_{t,t_m}  \nonumber
\eea
for all $N,M\geq 1$. 
For $N=0$, and $M=1$, 
\be
\label{eq:WTIb}
\HVEV \lim_{k^2\to0} \big( k^2 \Delta_\pi(k^2) \Big)= \HVEV
\ee
where
\bea
i H(;q,-q) &=& \int d^4x e^{iq\cdot x}\big< 0 \vert T\big[ \pi(x)\pi(0\big]\vert \big> \\
&\equiv& i \Delta_\pi(q^2)\nonumber
\eea
and $\Delta_\pi(q^2)$ is the pion propagator. 
The relation (\ref{eq:WTIb}) looks like the Goldstone Theorem,
and is indeed equivalent to it in this ungauged theory, 
where the masslessness of the NGB $\tilde\pi$ is equivalent to the masslessness of $\pi$,
as discussed in the body of the paper.

Equations (\ref{eq:WTIa}) and (\ref{eq:WTIb}) 
are the Ward-Takahashi Identities (WTI) of the theory.  
They are the fundamental identities upon which the arguments of this paper are based.
We can combine them to write (for $N\geq0, M\geq1$):
\bea
\label{eq:WTIcombined}
-\HVEV &&\left[i\Delta_\pi(0)\right]^{-1} H\left(p_1\cdots p_N; 0q_1\cdots q_M\right)\nonumber\\
&=& \sum_n^M H\left(p_1\cdots{\widehat {p_n}}\cdots p_N;p_n q_1 \cdots q_M\right)\quad \quad \\
&&-\sum_m^M H\left(p_1\cdots p_n q_m;q_1 \cdots {\widehat {q_m}}\cdots q_M\right)
 \nonumber
\nonumber
\eea

Equation (\ref{eq:WTIcombined}) is of the form
\bea
\label{eq:WTIform}
\HVEV (N+M+1)&-&{\rm point}~~{\rm function} \\
&& \propto \sum\left[(N+M)-{\rm point}~~{\rm functions}\right]\nonumber
\eea

B.W. Lee develops perturbation theory as an expansion in his $\lambda$,
the square-root  of his all-loop-orders renormalized 4-point coupling $\lambda^2$ 
(written less compactly $\lambda_\phi^2$ in the body of this paper).
Treating $\HVEV$ as $ {\cal O}(\lambda^{-1})$, 
Lee points out that equation  (\ref{eq:WTIcombined}), 
being of the form (\ref{eq:WTIform}),
``is satistied in each order of perturbation theory.''  
That is, if the $N+M+1$-point function on the LHS 
is computed in the l-loop approximation, 
that is to order $\lambda^{2(l-1)+N+M+1}$, 
and the $N+M$-point functions on the RHS are computed in the same l-loop approximation,
that is to order $\lambda^{2(l-1)+N+M}$, 
then the equation is identically satisfied.   

The off-mass-shell T-matrix for the $N$ scalar, $M$ pseudo-scalar process is obtained
from the connected amplitude 
$H(p_1\cdots p_N;q_1\cdots q_M)$ by ``amputating'' the propagators of the external lines:
\bea
&H(p_1\cdots p_N;q_1\cdots q_M) = \\
&\quad\quad \Pi_{n=1}^N \Big[ i\Delta_h(p_n^2)\Big] \Pi_{m=1}^M\Big[ i\Delta_\pi(q_m^2)\Big] T(p_1\cdots p_N;q_1\cdots q_M)
\nonumber
\eea

The off-shell 1-$(h,\pi)$ScalarParticle-Reducible  (1-$\phi$-R) connected amputated T-matrix elements  are expressed in terms of the all-loop-orders renormalized $h$ and $\pi$
propagators:
\bea
\label{Propagators}
i\Delta_h(p^2)&=& \int d^4x e^{i p \cdot x} \big<0\vert  T\big[  h(x)h(0)  \big]\vert 0\big>\quad \quad\\
i\delta_{ij}\Delta_\pi(q^2)&=& \int d^4x e^{i q \cdot x} \big<0\vert  T\big[\pi_i(x)\pi_j(0)\big]\vert 0\big>\nonumber
\eea
and 1-$(h,\pi)$ScalarParticle-Irreducible  (1-$\phi$-I) 
connected amputated Green's functions  $\Gamma_{N,M}(p_1\cdots p_N;q_1\cdots q_M)$,
which cannot be disconnected by cutting a $h$ or $\pi$ propagator line.
\bea
&T(p_1\cdots p_N;q_1\cdots q_M) =\\ 
&\quad \Gamma_{N,M}(p_1\cdots p_N;q_1\cdots q_M) + {\rm reducible~part}\,.\nonumber
\eea
$\Gamma_{N,M}(p_1\cdots p_N;q_1\cdots q_M)$ is the 1-$\phi$-I vertex for
$N$ $h$'s and $M$ $\pi$'s.  
The ``reducible part'' can be written in terms of irreducible vertices of lower order
and full propagators.  
Expressed in terms of the full propagators and the irreducible vertices, 
the T-matrix has a tree structure -- i.e. it can be represented by graphs without loops.

Because the T-Matrix contains only connected graphs, 
and our WTI concern only the axial-vector current,
and CP is converved,
\bea
\label{CPConservingSMatrixA}
\Gamma_{0,0}&=&\Gamma_{0,1}=\Gamma_{1,1}=0
\eea
In terms of connected amputated Green's functions, the propagators
\bea
\label{CPConservingSMatrixB}
\Gamma_{2,0}(p,-p;)&\equiv&\left[\Delta_h(p^2)\right]^{-1}\\
\Gamma_{0,2}(;q,-q)&\equiv&\left[\Delta_\pi(q^2)\right]^{-1}\nonumber
\eea

We now examine $\Gamma_{1,2}$.  By definition
\bea
H(p;0,-p) &=& \Big[i\Delta_h(p^2)\Big]\Big[i \Delta_\pi(0)\Big] \nonumber \\
&\times& \Big[i \Delta_\pi(p^2)\Big]  \Gamma_{1,2}(p;0,-p)
\eea
so with (\ref{eq:WTIcombined}) we have
\bea
\label{eq:Gamma12WTI}
\HVEV\Gamma_{1,2}(p;0,-p)&=&\left[\Delta_h(p^2)\right]^{-1} - \left[\Delta_\pi(p^2)\right]^{-1}\quad \quad \\
&=&\Gamma_{2,0}(p,-p;)-\Gamma_{0,2}(;p,-p)\,.\nonumber
\eea
We can easily verify that this holds to lowest order where 
$\Gamma_{1,2}(p;0,-p)=-2\lambda^2\HVEV$, 
$\left[\Delta_h(p^2)\right]^{-1}=p^2-m_h^2=p^2-2\lambda^2\HVEV^2$ and
$\left[\Delta_\pi(p^2)\right]^{-1}=p^2$.

To proceed further, we must express the WTI in terms of the connected amputated T-matrix rather than the
Green's functions.  
We can rewrite equation (\ref{eq:WTIcombined})
for $N\geq0$, $M\geq1$
\bea
\label{eq:TmatrixWTI}
&&\HVEV T(p_1\cdots p_N;0 q_1\cdots q_M) \\
&&\quad =
\sum_m^M i\Delta_h(q_m^2) \left[i\Delta_\pi(q_m^2)\right]^{-1} \nonumber \\ 
	&& \quad\quad \times T(p_1\cdots p_N q_m;q_1\cdots {\widehat {q_m}}\cdots q_M)\nonumber \\
&&\quad -\sum_n^N i\Delta_\pi(p_n^2) \left[i\Delta_h(p_n^2)\right]^{-1} \nonumber \\
	&& \quad\quad \times T(p_1\cdots  {\hat p}_t\cdots p_N ;q_1\cdots q_M p_t)\nonumber
\eea

A corollary of (\ref{eq:TmatrixWTI}) are Adler's self-consistency conditions for global $SU(2)_L\times U(1)_Y$:
\be
\label{eq:Adler}
\HVEV \lim_{q_1\to0}T(p_1\cdots p_N;q_1\cdots q_M)\vert_{
		p_1^2=\cdots=p_N^2=m_h^2}^{q_2^2=\cdots q_M^2=0} =0 \,,
\ee
which shows that, for $\HVEV \neq 0$, the T-matrix vanishes as one of the pion momenta goes to zero, provided
all the external particles are on the mass shell.
Equation (\ref{eq:Adler}) asserts the absence of infrared divergences in Goldstone mode!
``Individual Feynman diagrams are IR divergent, but the divergent parts must cancel 
in every order of perturbation theory.  
Furthermore, the amplitude must vanish in the soft-pion limit \cite{Lee1970}.''

In (\ref{eq:TmatrixWTI}), the zero-momentum pion in $T(p_1\cdots p_N;0 q_1\cdots q_M)$ can either
come off a ``branch'' (Lee's word for an external $\phi$ line) 
or off the ``body'' (our word) of the diagram.  
Let $T_1$ be the sum of the subset of the tree graphs belonging to 
$T(p_1\cdots p_N;0 q_1\cdots q_M)$ in which the zero-momentum pion comes
off a branch, as in Figure 1.  
The branch is either a $\pi$ branch (left-hand graph of Figure 1), with finite-momentum $q_m$, written
\be
\label{LHGraphFig1} 
i\Gamma_{1,2}(p_n;0,-p_n) i\Delta_\pi(p_n^2) 
T(p_1\cdots {\widehat {p_n}} \cdots p_N; p_n,q_1\cdots q_M) \nonumber
\ee
or a $h$ branch (right-hand graph of Figure 1), with momentum $q_m$, written
\be 
i\Gamma_{1,2}(q_m;0,-q_m) i\Delta_\pi(q_m^2) 
T(p_1\cdots p_N; q_1\cdots {\widehat {q_m}} \cdots  q_M) \nonumber
\ee
Forming $T_1$ from these, and using (\ref{eq:Gamma12WTI})
\begin{flalign}
\label{eq:T1WTI}
&&\HVEV T_1  
= \sum_m^M T(p_1\cdots  p_N q_m; q_1\cdots{\widehat {q_m}}  \cdots q_M) \nonumber \\
&&\times \Big(1- \left[i\Delta_\pi(q_m^2)\right]^{-1}\left[i\Delta_h(q_m^2)\right] \Big)\nonumber \\
&&-\sum_n^N T(p_1\cdots {\widehat {p_n}}  \cdots p_N; q_1\cdots q_M p_n)\nonumber\\
&&\times \Big(1- \left[i\Delta_h(p_n^2)\right]^{-1}\left[i\Delta_\pi(p_n^2)\right] \Big)  \quad \quad
\end{flalign}

Having accounted for $T_1$, we define 
\be
\label{DefinitionT2}
T_1+T_2 \equiv T(p_1\cdots  p_N;0 q_1\cdots q_M)
\ee
so that, combining (\ref{eq:TmatrixWTI},\ref{eq:T1WTI},\ref{DefinitionT2}), the WTI for $T_2$ are simply
\begin{flalign}
\label{eq:T2WTI}
	\HVEV &T_2(p_1\cdots  p_N;0 q_1\cdots q_M) &&\nonumber \\
	&= 
	\sum_m^M T(p_1\cdots  p_N q_m; q_1\cdots {\widehat {q_m}}  \cdots q_M) &&\nonumber \\
	&-
	\sum_n^N T(p_1\cdots {\widehat {p_n}}  \cdots p_N; q_1\cdots q_M p_n)&&
\end{flalign}
for $N\geq0$, $M\geq1$

These are identities for T-matrix elements.  How do they translate into
relations among the irreducible vertices? 
Lee shows that equation (\ref{eq:T2WTI})) is satisfied for $N\geq 0, M\geq1$ if 
\begin{flalign}
\label{eq:GammaWTI}
	\HVEV &\Gamma_{N,M+1}(p_1\cdots  p_N;0 q_1\cdots q_M) &&\\
	&= 
	\sum_m^M \Gamma_{N+1,M-1}(p_1\cdots  p_N q_m; q_1\cdots{\widehat {q_m}}  \cdots q_M) \nonumber&& \\
	&-
	\sum_n^N \Gamma_{N-1,M+1}(p_1\cdots{\widehat {p_n}} \cdots p_N; q_1\cdots q_M p_n)\,.\nonumber&& 
\end{flalign}
\vskip 1cm
\begin{figure}[!t]
\centering
\includegraphics[width=1\hsize,trim={0cm 5cm 0cm 5cm}]{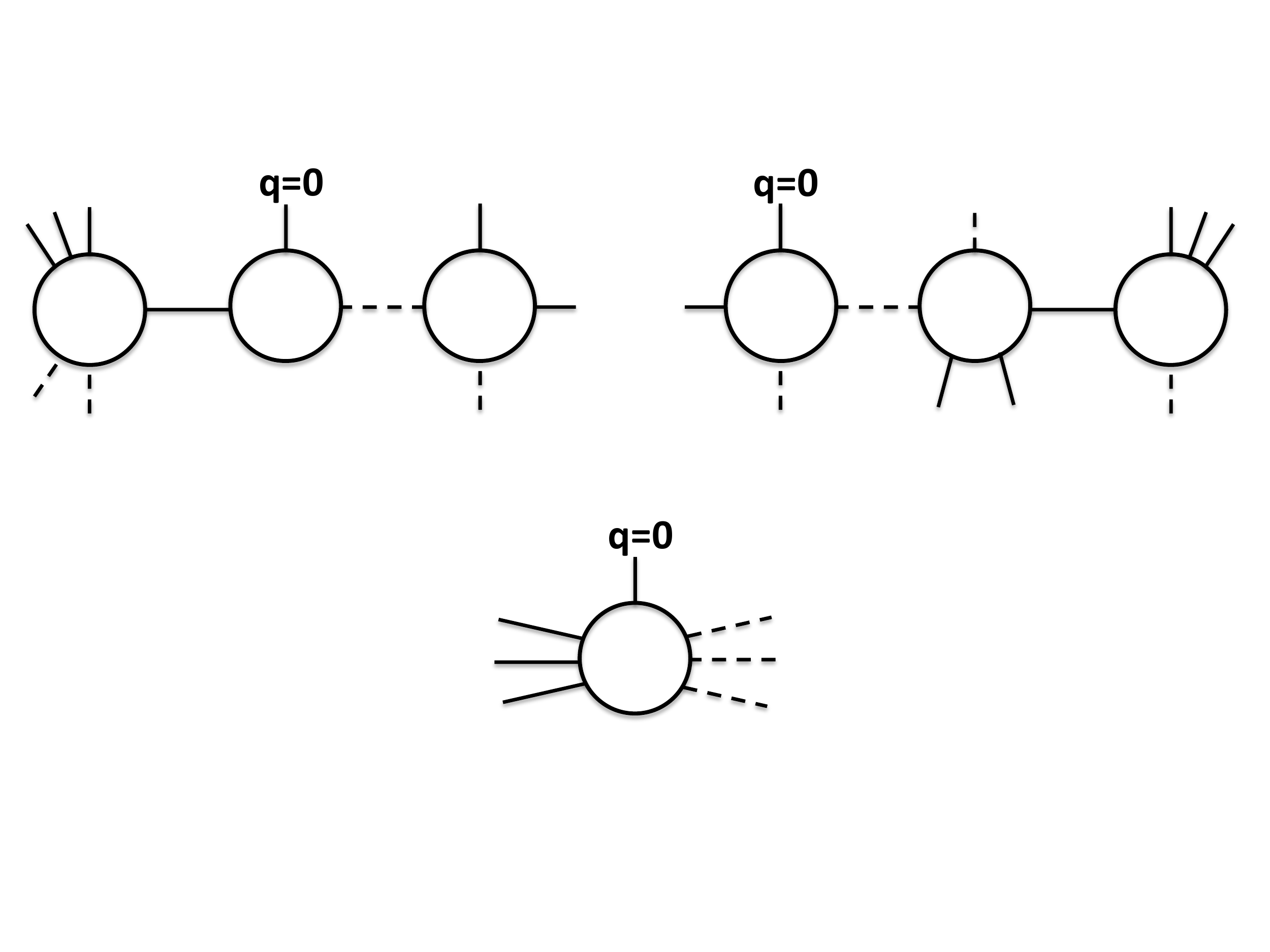}
\caption{
\label{fig:LeeFig11} 
Circles are 1-$\phi$-I $\Gamma_{n,m}$, solid lines $\vec \pi$, dashed lines $h$, with $n+m<N+M$.
One (zero-momentum) soft pion emerges in all possible ways from the  connected amputated Green's functions.
Fig. \ref{fig:LeeFig11}  is the $SU(3)_C\times SU(2)_L\times U(1)_Y$  $\GlnuDSM$ analogy of B.W. Lee's Figure 11 \cite{Lee1970}.
}
\end{figure}

The proof of (\ref{eq:GammaWTI}) 
is by induction on $N+M$, starting from $N=M=1$, 
which is just equation (\ref{eq:Gamma12WTI}).
Assume then that (\ref{eq:GammaWTI})  holds for $N+M<n+m$.
Let $N=n,M=m$. $T_2$ in (\ref{eq:GammaWTI}) contains
two classes of graphs, shown in Figure 2:

\begin{itemize}

\item {\bf Figure 2, top graphs} are reducible graphs in which the zero-momentum pion comes out
of an irreducible vertex.  However, this does not include graphs in which
the zero-momentum pion comes out of a three-prong irreducible vertex of which
two prongs are external lines, since those belong to $T_1$, not $T_2$.
  For the sum of the 1-$\phi$-R graphs, we may use (\ref{eq:GammaWTI}), 
for $N+M<n+m$, to show that the 1-$\phi$-I contributions from both sides of (\ref{eq:T2WTI})
are identical and cancel.  
This leaves only 1-$\phi$-I vertices on both sides of equation (\ref{eq:GammaWTI}),
giving us (\ref{eq:GammaWTI}) for $N=n,M=m$, as desired.

\item {\bf Figure 2, bottom graph} is 1-$\phi$-I, and already satisfies (\ref{eq:GammaWTI}).

\end{itemize}

Having proved equation (\ref{eq:GammaWTI}), we can now restore all the isospin
indices and display it in its full glory:
\begin{flalign}
\label{eq:GammaWTIfull}
	\HVEV &\Gamma_{N,M+1}^{t,t_1\cdots t_M}(p_1\cdots  p_N;0 q_1\cdots q_M) &&\\
	&= \sum_m^M \delta^{ t,t_m}
	\Gamma_{N+1,M-1}^{t_1\cdots {\widehat {t_m}} \cdots t_M}(p_1\cdots  p_N q_m; 
				q_1\cdots{\widehat {q_m}}  \cdots q_M) \nonumber && \\
	&-
	\sum_n^N \Gamma_{N-1,M+1}^{t_1\cdots t_M t}(p_1\cdots {\widehat {p_n}} \cdots p_N; 
				q_1\cdots q_M p_n)\,;\nonumber&& 
\end{flalign}
valid for all $N,M\geq0$, { and non-trivial for $M$ odd.}

As Lee emphasizes for the $SU(2)_L\times SU(2)_R$ symmetric theory,
{\em
"the identities (\ref{eq:GammaWTIfull}) are valid in 
any renormalizable theory in which the divergence [of the axial vector current vanishes] ...
and the $H,{\vec \pi}$ fields 
transform as the $[\half,\half]$ representation under chiral $SU(2)\times SU(2)$ transformations.
Whether ... other fields are included is irrelevant, so long as the chiral symmetry
is broken in a way to ensure the divergence remains zero.  When there are other
fields present, the irreducile vertices we have defined here may still
be reducible with respect to these [new] fields.''}. To Lee's statement, we add the strong constraint: {\bf as long as the 
new fields are massive!}

With the addition of massive Standard Model fermions to the $SU(2)_L\times SU(2)_R$ symmetric 
scalar theory, the symmetry is explicitly broken down to $SU(2)_L\times U(1)_Y$.

The addition of certain other new massive BSM particles that do not contribute to 
the divergence of the $SU(2)_L\times U(1)_Y$ current will again leave the form (\ref{eq:Adler},\ref{eq:GammaWTIfull})
of the WTI identities, and the Goldstone theorem, unchanged.  This is discussed more explicitly in section \ref{RenormalizationBSM}.

 \end{document}